\documentclass[aps,twocolumn,superscriptaddress,prb]{revtex4-2}

\usepackage[english]{babel}

\usepackage{amsmath,graphicx,csquotes,cases,physics,mathtools,amssymb,comment}
\usepackage[colorlinks=true, allcolors=black]{hyperref}
\usepackage{pgfplots}[compat=1.18]
\usepgfplotslibrary{fillbetween}
\pgfplotsset{compat = newest}
\usepgfplotslibrary{external} 
\newcommand{\bea}{\begin{eqnarray}}
\newcommand{\ea}{\end{eqnarray}}
\newcommand{\eea}{\end{eqnarray}}
\newcommand{\nn}{\nonumber\\}

\def\h{\hat}

\newcommand{\red}[1]{\textcolor{black}{#1}}
\newcommand{\change}[1]{\textcolor{black}{#1}}
\begin{document}
\title{Bound States at Semiconductor--Mott Insulator Interfaces}

\author{Jan Verlage} 

\affiliation{Fakult\"at f\"ur Physik and CENIDE, Universit\"at Duisburg-Essen,
  Lotharstra{\ss}e 1, 47057 Duisburg, Germany,}

\author{Peter Kratzer} 
\affiliation{Fakult\"at f\"ur Physik and CENIDE, Universit\"at Duisburg-Essen,
  Lotharstra{\ss}e 1, 47057 Duisburg, Germany,}

\date{\today}

\begin{abstract}
Utilizing the hierarchy of correlations in the context of a Fermi-Hubbard model, we deduce the presence of quasi-particle bound states at the interface between a Mott insulator and a semiconductor, as well as within a semiconductor--Mott--semiconductor heterostructure forming a quantum well. In the case of the solitary interface, the existence of bound states necessitates the presence of an additional perturbation with a minimal strength depending on the spin background of the Mott insulator.
Conversely, within the quantum well, this additional perturbation is still required to have bound states while  standing-wave solutions even exist in its absence.
\end{abstract}

\maketitle

\section{Introduction} 
 The continuous advancements in thin film growth techniques of oxides \cite{ohtomo2002artificial,yamada2004engineered} have sparked a surge of interest in 
 heterointerfaces between transition metal oxides \cite{hwang2012emergent,reyren2009anisotropy,mannhart2008two,bjaalie2014oxide}. Electrons within these complex oxides experience a strong on-site Coulomb repulsion \cite{zhou2005localized}, giving rise to intricate electronic correlations 
 while simpler oxides are band insulators or semiconductors. 
Interestingly, conducting interfaces between band-insulating oxide perovskites are well studied and understood, for example at the $\mathrm{LaAlO}_3/\mathrm{SrTiO}_3$ interface \cite{ohtomo2004high,thiel2006tunable}. 

The origin of the conducting layer 
is closely related to the avoidance of the so-called  'polar catastrophe', a mechanism broadly applicable to (001) interfaces  \cite{Maznichenko2020,maznichenko2019tunable,stemmer2014two,bristowe2014origin,hotta2007polar}. 
The polar heterojunction leads to charge transfer and subsequently to the formation and filling of a conduction band by the Ti $3d$ electrons \cite{Maznichenko2024}. 
Notably also the interface between the band insulator $\mathrm{KTaO}_3$ and Mott insulator $\mathrm{LaTiO}_3$ shows a conducting layer \cite{maryenko2023superconductivity,zou2015latio3}. In case of the (110) interface this cannot be traced back the 'polar catastrophe', instead the Mott insulating physics becomes relevant \cite{maryenko2023superconductivity}.

 Describing the strongly correlated nature of such Mott insulators necessitates specialized techniques such as strong-coupling perturbation theory  \cite{Pairault2000,Iskin2009}, dynamical mean-field theory \cite{Georges1996} or methods like DFT+U \cite{betancourt2017spin}, while semiconductors, on the other hand, can be 
 accurately described by established band-structure methods, e.g. many-body perturbation theory.
 
 The integration of these unlike materials into heterostructures holds the promise of unlocking novel applications \cite{https://doi.org/10.1002/admi.201900480,sharma2015mechanical,podkaminer2013creation,ji2020lateral,bjaalie2014oxide}.
 Because of this, it is crucial to have a comprehensive understanding of the electronic states at their interfaces. 
   While previous works have delved into the band lineup of heterointerfaces involving different Mott insulators and band insulators \cite{kitamura2015determination,nakamura2010interface} or calculated charge localization at specific Mott insulator interfaces like $\mathrm{LaAlO}_3/\mathrm{SrTiO}_3$ \cite{pentcheva2006charge}, $\mathrm{LaTiO}_3/\mathrm{KTaO}_3$ \cite{maznichenko2024fragile,Maznichenko2024} or $\mathrm{LaTiO}_3/\mathrm{SrTiO}_3$ \cite{santana2019electron,ishida2008origin,okamoto2004electronic}, this study 
   follows a general approach to deduce the existence of 
    bound states in such type of systems, independent of specific band lineups and materials.
 
 In navigating the complexities of these distinct material classes, we employ the hierarchy of correlations \cite{PhysRevA.82.063603,Queisser2019,PhysRevA.82.063603}, a framework that aligns both weakly and strongly interacting materials on equal footing while preserving spatial resolution \cite{verlage2024quasi}. This approach allows us to bridge the gap between the different description requirements of semiconductors and Mott insulators, offering a unified perspective on their behavior within heterostructures. \red{We 
 derive a Schrödinger-like equation for the quasi-particle wave functions 
 beyond the mean-field approximation.}
Within this approach, we will examine the different spin backgrounds of Mott insulators, the unpolarized background and the Mott-N\'eel state and their influence on charge bound states.
 
 At first, we introduce the Hubbard Hamiltonian used to model both the Mott insulator and the semiconductor as well as the hierarchy of correlations. 
 After this we calculate bound states at a single interface with an interface perturbation and give the minimal strength needed to support bound states for the different spin backgrounds. We continue with a heterostructure of an unpolarized Mott insulator stacked between two semiconductors with and without an interface perturbation.

\section{Hubbard Model and Hierarchy of Correlations}
To characterize bound states at the interfaces of systems exhibiting varying degrees of correlation strength, we employ the Hubbard model \cite{hubbard1963}, defined as follows:

\begin{equation}
\label{eq:FHMHamiltonian}
\hat{H} = -\frac{1}{Z}\sum_{\mu \nu s} T_{\mu \nu} c_{\mu s}^\dagger c_{\nu s} + \sum_\mu U_\mu \hat{n}_{\mu \uparrow} \hat{n}_{\mu \downarrow} + \sum_{\mu s} V_\mu \hat{n}_{\mu s}.
\end{equation}
Here, \(c_{\mu s}^\dagger\) and \(c_{\nu s}\) are the fermionic creation and annihilation operators, respectively, at sites \(\mu\) and \(\nu\). The corresponding number operators are denoted by \(\hat{n}_{\mu s}\), where the spin index \(s\) takes the values $\uparrow$ and $\downarrow$. The adjacency matrix \(T_{\mu \nu}\) encodes 
both the lattice structure and hopping strength, in this work taking the form \(T\) for nearest neighbors and zero otherwise. The coordination number \(Z\) counts the nearest neighbors.

\red{The parameter $U_\mu$ describes the on-site Coulomb interaction in the Fermi-Hubbard model, and it is non-zero only in the Mott insulator. In the following it sets the energy scale.}
Note that a possibly small $U$ in a weakly correlated semiconductor could be treated by a mean-field approximation and hence simply be absorbed in the magnitude of an effective $V_\mu$. This is similar to Fermi liquid theory \cite{landau1957theory,solovyev2017renormalized}. 

\red{In principle, both the Mott insulator and the weakly correlated layer have an on-site potential $V_\mu$, as the Hubbard bands do not necessarily need to be centered around $U/2$. But, as only the relative band alignment is crucial, we can set $V_{\mu \, \in \,\mathrm{Mott}} \equiv 0$ and describe the band alignment simply by the offset $V_{\mu \, \in \,\mathrm{semi}}-V_{\mu \, \in \,\mathrm{Mott}}\equiv V$. This incorporates both the material specific parameters of the band alignment and 
external electric fields into a single value for the band offset. For example, in the $\mathrm{LaTiO}_3/\mathrm{SrTiO_3}$ interface the specific band alignment depends on the number of grown layers. This would also be dealt with by the offset $V$.}

\red{Therefore,} the parameters \(U_\mu\) and \(V_\mu\) serve to differentiate between strongly correlated systems (\(U \neq 0\), \(V = 0\)) and weakly correlated systems (\(U = 0\), \(V \neq 0\)), as the on-site repulsion \(U_\mu\) and on-site potential \(V_\mu\) play crucial roles in characterizing the nature of electronic correlations and the relative band alignments within the system.

\subsection{Hierarchy of Correlations}
To approximate solutions for charge modes at the interface, we employ the hierarchy of correlations \cite{Queisser2014,Queisser2019,PhysRevA.82.063603} tailored for systems with a large coordination number \(Z \gg 1\). The reduced density matrices of two or more lattice sites are decomposed into correlated and on-site components. Specifically, for two lattice sites \(\mu\) and \(\nu\), the decomposition reads \(\hat{\rho}_{\mu \nu} = \hat{\rho}_\mu \hat{\rho}_\nu + \hat{\rho}_{\mu \nu}^{\text{corr}}\). Based on the large coordination number assumption $Z\gg1$, we may employ an expansion into powers 
of $1/Z$, where we find that higher-order correlators are successively
suppressed. 
The two-point correlator scales as $\hat\rho_{\mu\nu}^{\mathrm{corr}}=\mathcal{O}(Z^{-1})$,
while the three-point correlation is suppressed as 
$\hat\rho_{\mu\nu\lambda}^{\mathrm{corr}}=\mathcal{O}(Z^{-2})$, and so on.  This yields an iterative scheme to solve for the full density operator $\h \rho$.
More details are presented in App.~\ref{app:hierarchyofcorrelations}.

Similar to the idea of Hubbard X \cite{hubbard1965electron,ovchinnikov2004hubbard} or composite operators \cite{mancini2004hubbard}, we introduce quasi-particle operators as:
\bea
\hat c_{\mu s I}=\hat c_{\mu s}\hat n_{\mu\bar s}^I=
\left\{
\begin{array}{ccc}
	\hat c_{\mu s}(1-\hat n_{\mu\bar s}) & {\rm for} & I=0, 
	\\ 
	\hat c_{\mu s}\hat n_{\mu\bar s} & {\rm for} & I=1
\end{array}
\right.
\ea
 for doublons \(I=1\) and holes \(I=0\). 
 \red{These quasi-particles are the physical excitations within the Mott insulator on top of its half-filled background,
 and thus form a suitable starting point for describing 
 the physics.} 
 From these, we define the correlation functions $\langle \hat{c}_{\mu s I} \hat{c}_{\nu s J}\rangle^{\text{corr}}$ \cite{verlage2024quasi}. 
 \change{As we are interested in charge bound states, i.e., electrons and doublons/holons, we only take first order correlations into account. To this order, these are the only non-zero correlation functions. This leaves out the higher order correlations describing spin-fluctuations, particle number correlations and doublon-holon correlations, which would in second order act as source terms for the charge mode dynamics. Therefore, we treat the charge modes on top of a fixed mean-field background. This does not include any back-reaction effects of these modes onto the background. Incorporating this would effectively yield a renormalized hopping \cite{PhysRevA.89.033616_Queisser}, altering the dispersion relation. The neglected higher order correlations do not alter the dispersion and would solely act as source terms. For $Z \gg 1$ this is a valid approximation.}
 For the relevant part of the dynamics, this can further be simplified by a factorization \cite{navez2014quasi,Queisser2014}, 
which yields the doublon $p_\mu^1$ and holon $p_\mu^0$ amplitudes as  
\begin{equation}
\langle \hat{c}_{\mu s I} \hat{c}_{\nu s J}\rangle^{\text{corr}} = (p_\mu^I)^* p_\nu^J \, .
\end{equation} 
\red{In a sense, this is the same as writing the entries of the many-body density operator $\hat{\rho}$ in this doublon-holon basis as $\hat{\rho}_{IJ}=\left(p^I\right)^* p^J$ with the wave functions $p^I$.}
The two amplitudes can be grouped together using a spinor notation and 
governing equations for these (quasi) particles \cite{verlage2024quasi}
can be derived. 
\red{The interface in the systems breaks the translational invariance into this one direction, but leaves the other ones intact. Therefore, the parallel momentum $k^\|$ is still a conserved quantity. Hence, we decompose the wave functions $p_\mu^I=\sum_{k^\|} p_\mu^I(k^\|) e^{i k^\| \cdot x_\mu^\|}$ into their Fourier components. After this, $\mu$ is a scalar index counting the layers parallel to the interface, and for simplicity of notation we do not explicitly write down the momentum dependence.}
In a hyper-cubic lattice dependent on the parallel momentum $k^\|$ \red{the Schrödinger-like equations for the doublons $I=1$ and holons $I=0$} 
can be combined  using $U_\mu^I=I U_\mu$, 
\begin{equation}
	\label{difference}
\begin{aligned}
	\left(E-U_\mu^I-V_\mu\right)p_\mu^I + \langle\hat{n}_{\mu}^I\rangle^0\sum_J T_{\bf k}^\| p_\mu^J \\
	= -T\frac{\langle\hat{n}_{\mu}^I\rangle^0}{Z}\sum_J \left(p_{\mu-1}^J+p_{\mu+1}^J\right).
\end{aligned}
\end{equation}
\red{In this eq., $T_\mathbf{k}^{\|}=2 T/Z\sum_{i}\cos(k_{i}^\parallel)$ gives the kinetic energy contribution parallel to the interface. Because of the periodicity in this direction bands form.}
The expectation values $\langle\hat{n}_{\mu}^I\rangle^0$ are to be taken in the mean-field background $\hat{\rho}^0_\mu$ \red{and encode the spin background structure.} 
More details on the derivation are presented in App.~\ref{app:hierarchyforFermiHubbard}.
\red{Essentially, the hierarchy of correlations yields two coupled equations for the wave functions of doublons $p_\mu^1$ and holons $p_\mu^0$ on top of a half filled background discretized on the lattice. These can be solved for plane wave eigenmodes, evanescent solutions and boundary conditions, as known from the usual Schrödinger equation in quantum mechanics. We might formally write the system as}
\begin{equation}
\mathcal{H}_{\mu-1}\begin{bmatrix}
		p_{\mu-1}^{0} \\
		p_{\mu-1}^{1} 
	\end{bmatrix} + \mathcal{H}_{\mu+1}\begin{bmatrix}
		p_{\mu+1}^{0} \\
		p_{\mu+1}^{1} 
	\end{bmatrix}= \mathcal{H}_\mu \begin{bmatrix}
		p_{\mu}^{0} \\
		p_{\mu}^{1} 
	\end{bmatrix}.
\end{equation}
\red{This is the well-known form of 
a second-order differential equation in space, such as the Schrödinger
, Dirac or Klein-Gordon 
equation, for wave functions
discretized on a lattice.}

\subsection{Comparison to other Methods}
The hierarchy of correlations is based on a formal expansion into the inverse coordination number $1/Z$, not on a small parameter $T/U$ or $U/Z$ as in perturbation theory. This allows us to treat weakly and strongly correlated materials on the same level of theory.
\change{The dispersion relation within the strongly correlated Mott insulator might be calculated by means of other methods \cite{avella1998hubbard,hubbard1963electron,herrmann1997magnetism,roth1969electron,beenen1995superconductivity} like the Hubbard-I approximation, Roth’s two-pole scheme, the Second-Order Decoupling Approximation or Composite Operator Methods. 
Similar to the hierarchy of correlations, these methods rely on truncating an infinite series of expectation values or correlation functions. However, the hierarchy does have a small parameter $1/Z$, leading to a clear separation of leading and sub-leading contributions \cite{PhysRevA.100.053617,queisser2019boltzmann} controlling this truncation, while the other methods do not
(see, e.g., \cite{roth1969electron}, Eq. 56), making them inherently uncontrolled. 
Moreover, while the former approaches are best suited to thermal states, our framework naturally extends to dynamics (see \cite{queisser2023hierarchy} for the thermal case within the hierarchy). The hierarchy leads to non-perturbative results in $T/U$, as there are non-polynomial dependencies \cite{PhysRevA.89.033616_Queisser}. 
}

\change{An alternative method that has been used to calculate transport across Mott insulating layers \cite{okamoto2007nonequilibrium} and interface charge order \cite{okamoto2004spatial,Helmes2008} is 
Dynamical Mean-Field Theory (DMFT) \cite{Kotliar96}.}
\change{The guiding idea behind DMFT is different from our approach: }
\change{DMFT aims at approximating the self-energy of the system in order to find the system's Green's function, while the hierarchy aims at finding two- or multi-site correlation functions. The lifetime broadening related to self-energy is only included in higher orders.}
While in our approach the hopping scales as $1/Z$, DMFT uses a $1/\sqrt{Z}$ scaling. \change{As a consequence, the $Z\to \infty$ limit yields already a non-trivial and physically interesting result in DMFT, while in the hierarchy of correlations all correlations vanish in this limit.} This simplicity allows us to calculate the first-order $1/Z$ correlations on top of the mean-field background, and this includes the effects of the lattice structure and dimensionality. The hierarchy of correlations fully accounts for space-time dependence, particularly in higher dimensions \cite{queisser2023hierarchy,krutitsky2014propagation}, while DMFT \cite{Kotliar96} maps to an effective single lattice site. This single-site mapping leads to a purely local self-energy, neglecting all non-local correlations. DMFT is formally exact only in the limit of infinite dimensions, and corrections beyond this limit are generally uncontrolled. It is known to fail in low dimensions at low temperatures. In contrast, our approach incorporates non-local correlations, and the validity of the hierarchy of correlations is not dependent on the dimensionality of the system, but only on the number of neighbors. 
Moreover, the resulting equations are simple enough for analytic treatment, whereas DMFT 
is primarily numerical \cite{zujev2013induced,jiang2012density}.

\change{The hierarchy of correlations is expected to fail for low coordination numbers, as terms $1/Z^2$ cannot be neglected compared to $1/Z$ anymore, but even in these cases the results are still giving qualitative and quantitative agreement after including second order back-reaction effects and renormalizing the hopping strength, see Ref.~\cite{krutitsky2014propagation} for a comparison with exact diagonalization. Another situation where the method is expected to break down occurs when the two-point correlations become comparable in magnitude to the on-site expectation values. This typically happens near a phase transition, for instance during a quench from the Mott insulating to the superfluid phase in the Bose–Hubbard model. In such a regime, one would need to describe the evolution of the correlations on top of a newly determined mean-field background. Moreover, the hierarchy either works in the weakly or strongly interacting regime, but cannot work in the intermediate interacting strength relevant for, e.g., Kondo physics. DMFT, on the other hand, is capable of doing both things \cite{okamoto2004spatial}.}

To demonstrate the applicability, we assume a hyper-cubic lattice mostly for simplicity, though 
some relevant materials are cubic indeed.
Many perovskite Mott insulators, such as TiF$_3$ \cite{sheets2023mott} and LaTiO$_3$ \cite{maznichenko2024fragile} exhibit insulating behavior on three-dimensional cubic lattices with $Z=6$. Others occur on (quasi-)two-dimensional triangular lattices with the same coordination number \cite{PhysRevB.94.161105, tomeno2020triangular, PhysRevLett.99.256403}. Since the hierarchy depends on coordination number rather than dimensionality, our scaling applies equally to both cases. Second-order effects can also be included via a renormalized hopping parameter \cite{PhysRevA.89.033616_Queisser}, ensuring robustness across these systems. At present, such higher-order calculations exist only for homogeneous systems \cite{queisser2024back} and remain a subject for future work.

\subsection{Parameter Choice}
\red{Even though we describe the complex interactions in the Mott insulator using a simplified Fermi--Hubbard model, there are physically well-motivated choices for the parameters. A common strategy is to use DFT calculations as a complementary tool to obtain the band structure of the material of interest, then fit a tight-binding Hamiltonian to extract effective parameters. This approach generally yields a hopping-to-interaction ratio in the range
$0.05 < (T/Z)/U < 0.2$ \cite{okamoto2004spatial,okamoto2007nonequilibrium,popovic2005wedge,okamoto2006lattice,leonov2015metal}.}

\red{For instance, the DMFT study in Ref.~\cite{Okamoto2004} examined a heterostructure with a finite number of Mott-insulating layers embedded in an infinite band insulator. They used parameters $T = 0.3 \, \mathrm{eV}$ and $U = 4.8 \, \mathrm{eV}$, values extracted from experimental work on $\mathrm{SrTiO}_3/\mathrm{LaTiO}_3$ superlattices \cite{ohtomo2002artificial}, corresponding to $T/U = 0.0625$. In Ref.~\cite{okamoto2007nonequilibrium}, DMFT combined with the Keldysh formalism was employed to study a Mott insulator coupled to metallic leads, using $T/U = 0.066$ and a band offset of $V = U/2$. Similarly, Ref.~\cite{Helmes2008} investigated a Mott--metal interface with a band offset of approximately $V \approx 0.9U$.}

\red{Experimentally, band offsets have been reported in a range of systems: about $V = 0.05U$ for Ni--NiO--Ni junctions, $V = 0.3U$ for Ni--MnO--Ni junctions \cite{zhang2019dft+}, $V \approx 0.55U$ for a $\mathrm{LaVO}_3/\mathrm{SrTiO}_3$ interface \cite{stubinger2021hard}, and $V \approx 0.3U$ and $V \approx 0.25U$ for the conduction band offsets in $\mathrm{SrTiO}_3/\mathrm{SmTiO}_3$ and $\mathrm{SrTiO}_3/\mathrm{GdTiO}_3$, respectively \cite{bjaalie2016band}.}

\red{In the following, we will use either $T = 0.2U$ or $T=0.4U$ as the hopping strength. This yields an effective strength of $T/Z = 0.05U$ to $T/Z=0.1U$, right in the range of other studies.
We checked that varying the parameters within this range leads to 
only quantitative, but not qualitative changes, as the parameters remain within the (gapped) Mott-Hubbard regime.
}

\subsection{Unpolarized Mean-Field Background}
\label{subsec:unpolMott}
In the limit of strong correlation \(U \gg T\), the mean-field state within the strongly correlated Mott insulator ensures one particle per site, with additional virtual hopping processes suppressed by \(T^2/U^2\) \cite{zhou2005localized,iaconis2016kinetic}. 
This charge background has different manifestations as there are different spin orderings with the same charge configuration. At first, there is the unpolarized background:
\begin{equation} \hat{\rho}_\mu^0 = \frac{1}{2}\left(\ket{\uparrow}_\mu\bra{\uparrow} + \ket{\downarrow}_\mu\bra{\downarrow}\right). 
\end{equation}
This is realized by lattices with spin frustration or 
by a finite temperature destroying any spin ordering but too low to excite charges above the Mott gap $U$. 
\red{On top of this half-filled lattice the doublons (double occupations $\ket{\uparrow \downarrow}$) and holons (empty sites $\ket{0}$) are the physically relevant excitations for which we derived the Schrödinger-like equation.}

On the other hand, in the weakly correlated semiconductor region, where the quasi-particles are the real electrons, the on-site density matrix takes different forms for the valence and conduction bands:
\( \hat{\rho}_\mu^0 = \ket{\uparrow\downarrow}_\mu\!\bra{\uparrow\downarrow} \) for the valence band and \( \hat{\rho}_\mu^0 = \ket{0}_\mu\!\bra{0} \) for the conduction band.

Within each individual region, quasi-particles \red{eigenmodes} can be described using the \textit{ansatz} \(p_\mu^I = \alpha_I e^{i \kappa \mu} + \beta_I e^{-i \kappa \mu}\) \cite{verlage2024quasi}, with the proportionality $E p_\mu^0=(E-U)p_\mu^1$ in the Mott insulator. In the semiconducting sites, either one or the other is zero. \red{These describe the plane wave eigenmodes of the coupled doublons and holons in the Mott insulator and the electrons in the semiconductor, respectively.}
The corresponding wave numbers are given by
\begin{equation} \begin{aligned} \cos\kappa_\mathrm{semi} &= \frac{Z}{2T}\left[V - E - T_{\bf k}^\|\right], \\ \cos\kappa_\mathrm{Mott} &= \frac{Z}{2T}\left[\frac{E(U - E)}{E - U/2} - T_{\bf k}^\|\right].
\end{aligned} \end{equation}
These expressions provide a description of quasi-particle behavior within the semiconductor and Mott insulator 
regions, offering valuable insights into their wave-like properties in these 
correlated systems \cite{verlage2024quasi}.
In the following, 
we introduce the abbreviations:
\begin{equation}
\label{eq:unpolarized_abbreviations}
r_\pm=e^{\pm i \kappa_\mathrm{Mott}} \, ,  \quad s_\pm=e^{\pm i \kappa_\mathrm{semi}}
\end{equation}
to describe the Mott and semiconducting solutions, respectively.

\red{The quasi-particles (holons and doublons) show a wave like behavior inside the Mott bands and might be described by plane waves, similar to the electrons in the semiconducting band. The wave functions of these quasi-particles do not evolve independent of each other, but they are coupled. }

\subsection{Bi-Partite Mean-Field Background}
\label{subsec:bipartMott}
Secondly, there is the antiferromagnetic Mott-N\'eel state with its checkerboard structure:
\begin{equation}
\label{eq:rho0MN}
	\hat{\rho}_\mu^0= \begin{cases}
		\ket{\uparrow}\bra{\uparrow} & \mu \in A, \\
		\ket{\downarrow}\bra{\downarrow} & \mu \in B. \\
	\end{cases}
\end{equation}
\red{There are two different sublattices $A$ and $B$ with either spin-up or spin-down electrons. Without loss of generality, we use $\langle \h  n_{\mu_A \uparrow}\rangle=1$, $\langle \h  n_{\mu_B \uparrow}\rangle=0$.} 
In this spin configuration, the unit cell has double the size compared to the unpolarized background. This leads to backfolding in the Brillouin zone, giving two solutions: an even and an odd one. Moreover, the bi-partite structure will imprint on the weakly interacting sites.

\red{In order to capture the quasi-particles and their eigenmodes on this bi-partite lattice structure,} we introduce another index $A$ or $B$ for the respective sublattice. In the Mott-N\'eel state, the coupled equations for doublons and holons thus read
\begin{equation}
\label{bipartite-spinors}
	\begin{aligned}
		E p_\mu^{0_A}&= - \left[T_\mathbf{k}^\| p_\mu^{1_B}+\frac{T}{Z}\left( p_{\mu+1}^{1_B}+p_{\mu-1}^{1_B}\right)\right], \\
			\left(E-U\right) p_\mu^{1_B}&= - \left[T_\mathbf{k}^\| p_\mu^{0_A}+\frac{T}{Z}\left( p_{\mu+1}^{0_A}+p_{\mu-1}^{0_A}\right)\right].
	\end{aligned}
\end{equation}
\red{As in the unpolarized state, doublons and holons are not independent of each other, but they are coupled.}
\red{This leads to a} proportionality between the two quasi-particles $p_\mu^{1_B}=\beta p_\mu^{0_A}$ with $\beta= \pm \sqrt{\frac{E}{E-U}}$.
 With the \textit{ansatz} $p_\mu^{1_B}=\mathcal{B}\kappa^\mu$ and $p_\mu^{0_A}=\mathcal{A}\kappa^\mu$, \red{$\mathcal{A}$ and $\mathcal{B}$ being the wave function amplitudes,} we find the eigenmodes
\begin{equation}
	\begin{aligned}
		\kappa_{1,2}=&-\frac{Z}{2T} \left(\sqrt{E(E-U)}+T_\mathbf{k}^\| \right)\\&\pm\sqrt{\left[\frac{Z}{2T} \left(\sqrt{E(E-U)}+T_\mathbf{k}^\| \right) \right]^2-1},\\
		\kappa_{3,4}=&+\frac{Z}{2T} \left(\sqrt{E(E-U)}-T_\mathbf{k}^\| \right)\\&\pm\sqrt{\left[\frac{Z}{2T} \left(\sqrt{E(E-U)}-T_\mathbf{k}^\| \right) \right]^2-1}\\
	\end{aligned}
\end{equation}
with $\kappa_1 \kappa_2=\kappa_3 \kappa_4=1$. \red{Because of the backfolding there are four eigenmodes. Within the Hubbard bands, these obey $|\kappa_i|=1$, such that we have plane waves again. This can be seen from the identity $x\pm i \sqrt{1-x^2}=e^{\pm i \arccos(x)}$, which defines the wave numbers for the two different types of plane wave eigenmodes $\kappa_{1,2}$ and $\kappa_{3,4}$.
Within the Mott bands, they read:}
\begin{equation}
\begin{aligned}
    \cos(x_{1,2}) &= \frac{Z}{2T} \left(\sqrt{E(E-U)}+T_\mathbf{k}^\| \right), \\
        \cos(x_{3,4}) &= \frac{Z}{2T} \left(\sqrt{E(E-U)}-T_\mathbf{k}^\| \right). 
\end{aligned}
\end{equation}
\red{Outside the Mott bands, they define decaying solutions.}
The proportionality between them reads 
\begin{equation}
	\label{eq:BiAieq}
	\mathcal{B}_i = \frac{1}{U-E}\left[T_\mathbf{k}^\|+\frac{T}{Z}\left(\kappa_i + \kappa_i^{-1}\right) \right] \mathcal{A}_i,
\end{equation}
which again fixes $\beta$. $\kappa_1$ and $\kappa_2$ 
belong to 'even' solutions, i.e. having the same sign on neighboring sites of both sublattices, whereas $\kappa_3$ and $\kappa_4$ belong to 'odd' solutions defined by a sign change between sublattices.

Without loss of generality, we can extend the bi-partite structure to the semiconducting half-space, now setting 
$p_\mu^{0_A}=\pm \alpha p_\mu^{0_B}$ in the conduction band case. \red{The eigenmodes on the two sublattices are not independent of each other.} Together with the \textit{ansatz} $\lambda^\mu$, we find the eigenmodes for $\alpha=+1$ as:
\begin{equation}
	 \lambda_\pm = -\frac{Z}{2T} \left(E-V+T_\mathbf{k}^\| \right) \pm \sqrt{\left[ \frac{Z}{2T} \left(E-V+T_\mathbf{k}^\| \right)\right]^2-1}.
\end{equation}
This is again an 'even' solution.
Similar for $\alpha=-1$, we find with $p_\mu^{0_A}=\rho^\mu$ 
\begin{equation}
	\rho_\pm = +\frac{Z}{2T} \left(E-V-T_\mathbf{k}^\| \right) \pm \sqrt{\left[ \frac{Z}{2T} \left(E-V-T_\mathbf{k}^\| \right)\right]^2-1}.
\end{equation}
belonging to the 'odd' solution. \red{Inside the semiconducting band these are plane waves $e^{i k}$ with wave numbers 
defined as solution of 
$\cos(k)=\frac{Z}{2T}(E-V+T^\|_\mathbf{k})$ and $\cos(k)=\frac{Z}{2T}(E-V-T^\|_\mathbf{k})$, respectively.}

\red{As in the unpolarized case, the quasi-particles on the bi-partite lattice show a wave-like behavior within the bands. They are coupled, and the wave functions on the different sublattices are proportional to each other with an energy dependent factor. Moreover, there are two different solutions, even and odd ones, with the same sign or a sign switch between the wave functions on their respective sublattice.}

\section{Unpolarized Mott-Semiconductor interface}
\begin{figure}
	\centering
	\begin{tikzpicture} 
		\begin{axis}[
			xmin = -0.5, xmax =1.5,
			ymin = -0.1, ymax =4.5,
			xtick distance = .5,
			ytick distance = 1,
			restrict y to domain=-10:10,
			minor tick num = 1,
			major grid style = {lightgray},
			minor grid style = {lightgray!25},
			width = \linewidth,
			height = 0.7\linewidth,
			xlabel = {$E/U$},
			ylabel = {$D(E)$ [a.u]},
			legend cell align = {right},
			cycle list name=black white,
			legend style={at={(.95,.85)},anchor=east},
                yticklabel=\empty
			]

			\addplot[
            blue,
			dashed, line width=1.5pt
			] table[x index = 0, y index = 1] 
            {pics/dos_data_mott.txt};

			\addplot[
            red, line width=1.5pt
			]  table[x index = 0, y index = 1] {pics/dos_data_semi.txt};


			\draw[line width=1.pt, dashed] (0.190863,-0.2) -- (0.190863,4.5);
                \draw[line width=1.pt, dashed] (-0.20863,-0.2) -- (-0.20863,4.5);
			
			\legend{Mott, Semiconductor}
		\end{axis}
		
	\end{tikzpicture}
	\caption{\red{Density of states $D(E)$ of the Mott insulator (blue dashed) and the semiconductor (red solid) together with the delta-peaks of the bound states shown as the black dashed vertical lines.}}
	\label{fig:schematics_defeq}
\end{figure}

First, our objective is a single unpolarized Mott-semiconductor interface. We consider a hypercubic lattice where the two half-spaces correspond to a Mott insulator with site index \(\mu < 0\) and a semiconductor with \(\mu \geq 0\). 
In addition to the infinitely extended continuum states \cite{verlage2024quasi}, bound states, characterized by 
their spatial decay away from the interface, may occur.
For the \red{quasi-particle wave functions} $r_a$ and $s_b$ in our \textit{ansatz}, this implies
\begin{equation}
\label{eq:ansatz}
p_{\mu}^{0}=
\begin{cases}
	A_{a}\left(r_{a}\right)^{\mu}  &\mu<0 \text { with }\left|r_{a}\right|>1 \text{ in Mott}, \\
	B_{b}\left(s_{b}\right)^{\mu} & \mu \geq 0 \text { with }\left|s_{b}\right|<1 \text{ in semi.},
\end{cases}
\end{equation}
where the indices $a,b$ stand for the $+$ or $-$ sign \red{in Eq.~\ref{eq:unpolarized_abbreviations}}. \red{Outside their respective bands, both $\kappa_\mathrm{Mott}$ and $\kappa_\mathrm{semi}$ are purely imaginary, such that the holon wave function $p_\mu^0$ decays away from the interface to both sites. The bound state energy finally decides if the physically sound solution is comprised of the $+$ or $-$ sign solution.}

While we keep the model parameters $T_{\mu \nu}, U_\mu$ and $V_\mu$ constant inside their respective regions, our 
analysis shows that an additional on-site \red{perturbation} \(\Delta V \neq 0\) at the semiconductor interface site 
\(\mu=0\) is a pre-requisite for obtaining interface states.
This interface perturbation arises due to local modifications in the epitaxial interfaces of different materials. It might be introduced by a structural relaxation or deformation \cite{wong2010metallicity,okamoto2006lattice,schoofs2013carrier}, by the tilting of the perovskite octahedra as in $\mathrm{LaTiO}_3/\mathrm{KTaO}_3$ \cite{Maznichenko2024} or other local modifications \cite{fister2014octahedral,maurice2006electronic}.

By considering the relation between doublons \(p_\mu^1\) and holons \(p_\mu^0\) in the Mott and semiconductor regions, we can establish boundary conditions at the interface using Eq.~\ref{difference}. These relate the amplitudes with the interface perturbation. 
After some algebra (see App.~\ref{App:single_interface} for details), the defining equation for bound states at the single unpolarized interface reads:
\begin{equation}
\label{eq:delta_v_def}
\frac{1}{r_a} - \frac{1}{s_b} = \frac{\Delta V Z}{T}.
\end{equation}
It is important to note that this condition lacks a solution in the absence of an interface perturbation (see App.~\ref{App:single_interface_defEq}). \red{Without it, the interface only supports interface resonances, decaying to one site while being a plane wave in the other half-space.} \red{Fig.~\ref{fig:schematics_defeq} shows the density of states of this system as the bands of the Mott and semiconductor together  with the delta-like bound state energies.} It looks similar to the Newns-Anderson model describing chemisorption or localized magnetic states in metals \cite{newns1969self,anderson1961localized}.
%
Fig.~\ref{fig:example_dV} shows one solution to this equation.
As expected for a bound state, the probability distribution exhibits localization at the interface -- on the lattice site with the interface perturbation -- and displays distinct decay constants towards both the 
Mott insulator and the semiconductor regions.

\red{We find bound states solutions for attractive and repulsive interface perturbations. This is a distinction from the usual quantum mechanics case with a delta potential, in which only the attractive case has bound state solutions.}

\begin{figure}
	\centering
	\begin{tikzpicture} 
		\begin{axis}[
			xmin = -20, xmax =20,
			ymin = -1, ymax =1,
			xtick distance = 5,
			ytick distance = .5,
			restrict y to domain=-10:10,
			minor tick num = 1,
			major grid style = {lightgray},
			minor grid style = {lightgray!25},
			width = \linewidth,
			height = 0.7\linewidth,
			xlabel = {site index $\mu$},
			ylabel = {holon wave function $p_\mu^0$},
			legend cell align = {right},
			cycle list name=black white,
			legend style={at={(.95,.2)},anchor=east}
			]

			\addplot[
			dashed, red, line width=1.5pt
			] table[x index = 0, y index = 1] {pics/plot_example.dat};

			\addplot[line width=1.5pt
			]  table[x index = 0, y index = 2] {pics/plot_example.dat};

			\draw[dotted] (0,-2)--(0,2);
			
			\node[draw] at (-5,0.5) {Mott};
						\node[draw] at (10,.5) {Semiconductor};		
			\legend{$p_\mu^0$,$|p_\mu^0|^2$}

		\end{axis}
		
	\end{tikzpicture}
	\caption{Bound state \red{for the holon wave function }$p_\mu^0$ and quasi particle probability distribution $|p_\mu^0|^2$  at the interface for the unpolarized background with $E=1.16U$. The parameters are $V=1.1U$, $T=0.4U$ and $\Delta V=0.2U$.}
	\label{fig:example_dV}
\end{figure}

\subsection{Minimal Interface Perturbation}
From an experimental standpoint, the selection of materials for the Mott insulator and the semiconductor will determine the interface perturbation $\Delta V$, \red{the hopping strength $T$ as well as the band offset $V$}. Consequently, it is crucial to comprehend the requisite strength of the interface perturbation to establish a bound state. 
\red{Without this, the interface only supports interface resonances or, in case the band overlap energetically, quasi-particle wave functions covering both half-spaces.}
Interestingly, such a state is discovered to exist irrespective of whether $\Delta V$ is attractive or repulsive, provided it surpasses a specific threshold.
\red{While experimentally the choice of materials fixes 
the parameters, in our theoretical study 
it is instructive to tune the band offset $V$. As bound states have an energy outside of the Mott and semiconducting band, the respective band edges serve as a natural boundary for the allowed energies. For this, we need to distinguish between an attractive and repulsive perturbation.} 

\red{In the \(\Delta V > 0\) case, bound states are comprised of \(r_a = r_+\) and \(s_b = s_-\).}
To satisfy the condition \(|s_b| = |s_-| < 1\) any valid solution requires an energy 
\(E > B = 2 T/Z - T_\mathbf{k}^\|  + V \), ensuring that it 
lies energetically above the semiconducting band.
To fulfill \(|r_+| > 1\), the 
upper band edges of the lower \(A_L\) and 
of the upper \(A_H\) Hubbard band are relevant:
\begin{equation}
\begin{aligned}
	A_L &= \frac{1}{2} \left[-\sqrt{\left(T^H_\mathbf{k}\right)^2 + U^2} + T^H_\mathbf{k} + 2U \right], \\
	A_H &= \frac{1}{2} \left[\sqrt{\left(T^H_\mathbf{k}\right)^2 + U^2} + T^H_\mathbf{k} + 2U \right],
\end{aligned}
\end{equation}
where \(T^H_\mathbf{k} =  \frac{2T}{Z} - T_\mathbf{k}^\|\).

Depending on the model parameters \(V\) and \(U\), four scenarios of band alignment arise.
Firstly, the semiconducting band edge might be energetically higher than the Mott bands $B>A_H$, or equivalently, \(V > U\) for \(k^\| = 0\). In this case, the threshold for the existence of bound states is given by Eq.~\ref{eq:delta_v_def} at \(E = B\):
\begin{equation}
\Delta V_\mathrm{min,B} = \frac{T}{Z}\left(\frac{1}{r_+(E=B)} - \frac{1}{s_-(E=B)}\right).
\end{equation}
In the limit \(V \gg U\) this approaches \(\Delta V_\mathrm{min,B} \to \frac{T}{Z}\). In this case, the required interface perturbation is small, \(\Delta V \ll U\).

Secondly, for a semiconducting band edge situated between the center of the Mott gap and the upper Hubbard band $U/2 < B < A_H$, the \textit{ansatz} implies \(E > A_H\). \red{This is the case in Ref.~\cite{Helmes2008}.} Consequently, Eq.~\ref{eq:delta_v_def} at \(E = A_H\) yields:
\begin{equation}
\Delta V_\mathrm{min,A_H} = \frac{T}{Z}\left(\frac{1}{r_+(E=A_H)} - \frac{1}{s_-(E=A_H)}\right).
\end{equation}
For strong Coulomb interaction compared to the hopping, this simplifies to \(\Delta V_\mathrm{min,A_H} = \frac{1}{2}T_\mathbf{k}^\| + U - V + \mathcal{O}(T^2)\).

Thirdly, if the semiconducting band edge is between the lower Hubbard band and the middle of the Mott band gap \(A_L < B < U/2\), any bound state has \(B < E < U/2\). The threshold is found as \(\Delta V_\mathrm{min,B}\). \red{This is the case in Ref.~\cite{bjaalie2016band}.}

Fourthly, the semiconducting band edge might be energetically lower than the Hubbard bands \(B < A_L\)), which requieres \(A_L < E\) and \(A_L < E < U/2\) for bound states. Thus, Eq.~\ref{eq:delta_v_def} at \(E = A_L\) gives:
\begin{equation}
\Delta V_\mathrm{min,A_L} = \frac{T}{Z}\left(\frac{1}{r_+(E=A_L)} - \frac{1}{s_-(E=A_L)}\right).
\end{equation}
To first order in the hopping, this reads \(\Delta V_\mathrm{min,A_L} = \frac{1}{2}T_\mathbf{k}^\| - V + \mathcal{O}(T^2)\).
For the full expressions in terms of the parameters \(V\), \(U\), \(T\), and \(T_\mathbf{k}^\|\) see App.~\ref{App:single_interface_minimalpot}.

The required strength of the interface perturbation \(\Delta V_\mathrm{min}\) depends on the band alignment. It is depicted in Fig.~\ref{fig:dV_min_v} (sold line) as a function of this band alignment, parametrized by the potential \red{offset }$V$ for given hopping $T$. The approximate dependence of $T$, $U$, and $V$ is given. It decreases linearly with decreasing distance $V$ from the lower Hubbard band; is nearly constant in the lower half of the Mott gap; it goes up at $V=U/2$ and decreases from there linearly; before being constant again. \red{For an existing interface, the parameter combination is given by the materials such that this system would be a single point on either the solid or dashed line (two examples are marked). By a gate voltage applied to only one half-space, the band offset $V$ may be shifted, and the minimally needed strength moves accordingly.
}

\begin{figure}
	\centering
	\begin{tikzpicture} 
		\begin{axis}[
			xmin = -0.5, xmax =2,
			ymin = -1, ymax = 2.4,
			xtick distance = .5,
			ytick distance = .5,
			restrict y to domain=-10:10,
			minor tick num = 1,
			major grid style = {lightgray},
			minor grid style = {lightgray!25},
			width = \linewidth,
			height = 0.7\linewidth,
			xlabel = {potential offset $V/U$},
			ylabel = {$\Delta V_\mathrm{min}/U$},
			legend cell align = {right},
			cycle list name=black white,
			legend style={at={(0.99,0.75)},anchor=east}
			]

			\addplot[
            line width=2pt
			] table[x index = 0, y index = 1] {pics/dv_min_T04kp0_overV.dat};
			

				\addplot[
			dashed, red, line width=2pt
			] table[x index = 0, y index = 1] {pics/min_dV_negdV.dat};


						\legend{ $\Delta V_\mathrm{min}/U>0$, $\Delta V_\mathrm{min}/U<0$}
			
												\node at (1.5,.5) {$T/Z$};
												\node at (.25,.5) {$T/Z$};
											\node[align=left] at (.75,1.){$U-V$\\$+\frac{T_\mathbf{k}^\|}{2}$}; 
											\node at (-0.25,0.9) {$\frac{T_\mathbf{k}^\|}{2}  - V$};

                    \draw[blue] (0.55,-2) -- (0.55,2.2) node[right] {$\mathrm{STO}/\mathrm{LaVO}_3$ };

                    \draw[orange] (0.3,-2) -- (0.3,2.2) node[left] { $\mathrm{STO}/\mathrm{SmTiO}_3$ };
		\end{axis}
		
	\end{tikzpicture}
	\caption{Minimally needed interface perturbation $\Delta V_\mathrm{min}$ in the positive (solid) and negative (dashed) case as a function of the on-site potential offset $V$. The hopping strength $T=0.4U$ is fixed. The \red{annotations} give the approximate formula in the $\Delta V>0$ case. \red{The vertical lines give the band offsets for $\mathrm{SrTiO}_3$ interface with $\mathrm{SmTiO}_3$ and $\mathrm{LaVO}_3$, respectively.} }
	\label{fig:dV_min_v}
\end{figure}

Similar considerations as above might also be done for 
a negative interface perturbation \(\Delta V < 0\), corresponding to
\(r_a = r_-\) and \(s_b = s_+\). 
In this case, any valid solution requires an energy below the semiconducting band edge 
\(E < B_L = -2 T/Z - T_\mathbf{k}^\| + V \), where both lower band edges in the Mott region, 
\(A_{LL}\) and \(A_{LH}\), are relevant. With \(T^L_\mathbf{k} = 2T/Z + T_\mathbf{k}^\|\) they read 
\begin{equation}
\begin{aligned}
	A_{LL} &= -\frac{1}{2}\left[T_\mathbf{k}^L - U + \sqrt{(T_\mathbf{k}^L)^2 + U^2}\right], \\
	A_{LH} &= -\frac{1}{2}\left[T_\mathbf{k}^L - U - \sqrt{(T_\mathbf{k}^L)^2 + U^2}\right].
\end{aligned}
\end{equation}

The minimum required interface perturbation can be calculated using the same approach as before,  
and is shown in Fig.~\ref{fig:dV_min_v} (dashed line). 
The solutions for attractive and repulsive $\Delta V$ seem to be related (at least approximately) via an inflection point inside the Hubbard gap. 

\subsection{Decay Constant}
The decay on both sides, \(e^{- \kappa |\mu|}\), is characterized by 
an inverse length scale: 
\begin{equation}
\begin{aligned}
	\kappa_\mathrm{Mott} &= \mathrm{arcosh}\left(\abs{\frac{Z}{2T}\left[\frac{E(U-E)}{E-U/2}-T_{\bf k}^\|\right]}\right), \\
	\kappa_\mathrm{semi} &= \mathrm{arcosh}\left(\abs{\frac{Z}{2T}\left[V-E-T_{\bf k}^\|\right]}\right).
\end{aligned}
\end{equation}
The decay constant for the semiconducting side is illustrated in Fig.~\ref{fig:deconst_semi_v}. 
Notably for small \red{but realistic} \(T\), there exists a range of interface perturbations without any solutions (e.g., for \(T=0.2U\), from \(\Delta V/U \approx 0.5 - 1\)) \red{for a band offset close to the lower Hubbard band $V=0.2U$, which is close to the measured ones for $\mathrm{SrTiO}_3/\mathrm{SmTiO}_3$ and $\mathrm{SrTiO}_3/\mathrm{GdTiO}_3$ \cite{bjaalie2016band}.}

\begin{figure}
	\centering
	\begin{tikzpicture} 
		\begin{axis}[
			xmin = 0, xmax =2.5,
			ymin = 0, ymax = 4,
			xtick distance = 1,
			ytick distance = 1,
			restrict y to domain=-10:10,
			minor tick num = 1,
			major grid style = {lightgray},
			minor grid style = {lightgray!25},
			width = \linewidth,
			height = 0.7\linewidth,
			xlabel = {$\Delta V/U$},
			ylabel = {$\kappa_\mathrm{semi}$},
			legend cell align = {right},
			cycle list={%
			{mark=asterisk},
			{mark=o},
			{mark=square}
		},
				legend style={at={(.95,.3)},anchor=east}
			]
			
			
				\addplot+[
			thin, mark size=2pt, only marks, mark repeat=3,mark phase=1
			] table[x index = 0, y index = 1 ] {pics/decay_semi_fix_1p102.dat};
			
			
			
				\addplot+[
				thin,orange, mark size=2pt,mark options={scale=1, fill=orange},only marks,mark repeat=3,mark phase=1
				] table[x index = 0, y index = 1 ] {pics/decay_const_one_int_v1p5_T0p20p80p3_semi_denser.dat};

				\addplot+[
			thin, red, mark size=2pt,mark options={ fill=red}, only marks,mark repeat=3,mark phase=1
			] table[x index = 0, y index = 1 ] {pics/decay_semi_fix_0202.dat};
			
			

			
			\legend{$V=1.1U$,$1.5U$,$0.2U$}

		\end{axis}
		
	\end{tikzpicture}
	\caption{The decay constant $\kappa_{\text{semi}}$ in the unpolarized semiconducting half-space \red{is shown as a function of the interface perturbation for different band offsets $V$.} The hopping strength is fixed at $T=0.2U$. \red{Note that gap in the $0.2 U$ data.}}
	\label{fig:deconst_semi_v}
\end{figure}

\begin{figure}
	\centering
	\begin{tikzpicture} 
		\begin{axis}[
			xmin = 0, xmax =4,
			ymin = 0, ymax = 6,
			xtick distance = 1,
			ytick distance = 1,
			restrict y to domain=-10:10,
			minor tick num = 1,
			major grid style = {lightgray},
			minor grid style = {lightgray!25},
			width = \linewidth,
			height = 0.7\linewidth,
			xlabel = {$\Delta V/U$},
			ylabel = {$\kappa_\mathrm{Mott}$},
			legend cell align = {right},
			legend style={at={(.95,.3)},anchor=east},
			cycle list={%
			{mark=*},
			{mark=triangle*},
						{mark=square*}
			}
			]

				\foreach \x in {1}
			\addplot+[
			thin, mark size=2pt,  only marks,mark repeat=3,mark phase=1
			] table[x index = 0, y index = \x ] {pics/decay_const_one_int_v1p1_T0p20p80p3_mott_denser.dat};
			
			\foreach \x in {1}
			\addplot+[
			thin, blue, mark size=2pt,mark options={ fill=blue},only marks,mark repeat=3,mark phase=1
			] table[x index = 0, y index = \x ] {pics/decay_const_one_int_v1p5_T0p20p80p3_mott_denser.dat};

			
				\addplot+[
			thin, red, mark size=2pt,mark options={ fill=red}, only marks,mark repeat=3,mark phase=1
			] table[x index = 0, y index = 1 ] {pics/decay_mott_fix_0202.dat};
			
			
			

			\legend{$V=1.1U$,$1.5U$,$0.2U$}

		\end{axis}
		
	\end{tikzpicture}
	\caption{The decay constant $\kappa_{\text{Mott}}$ in the unpolarized Mott half-space is shown for different band offsets $V$ as a function of the interface perturbation $\Delta V$. The hopping strength is fixed at $T=0.2U$.}
	\label{fig:deconst_mott_v}
\end{figure}
Fig.~\ref{fig:deconst_mott_v} illustrates the same information for the Mott side. 
Again, 
for \red{realistic} hopping parameters 
there is a range of $\Delta V$ values where no bound-state solution exists. 
In contrast to the semiconducting side, the decay constant in the Mott insulator 
may display non-monotonic behavior: 
When the \red{band offset} \(V\) is close to the lower Hubbard band, a discontinuity occurs 
where \(\kappa_\mathrm{Mott}\) sharply rises to a large value before abruptly decreasing close to zero. 
Increasing the interface perturbation for such a semiconducting band causes the energy of the solution to transition from being between the Hubbard bands to residing above the upper band, resulting in the observed discontinuity.

The rate of decay, whether it occurs more rapidly in the Mott or the semiconducting side, depends on both the \red{band offset} and the hopping strength.
Fig.~\ref{fig:deconst_mott_semi_vgl} displays both decay constants for a \red{band offset} close to the lower Hubbard band and another above the upper one, both for a small \red{but realistic} hopping strength. In the case of the larger on-site potential, the decay constant in the Mott is greater than in the semiconductor, indicating faster decay in the Mott. However, for the smaller on-site potential near the lower Hubbard band, there is a reversal as the interface perturbation increases. For small interface perturbations, the decay is faster in the Mott, but as the interface perturbation reaches a certain magnitude, the decay becomes slower than in the semiconductor.

\begin{figure}
	\centering
	\begin{tikzpicture} 
		\begin{axis}[
			xmin = 0, xmax =4.1,
			ymin = 0, ymax = 7,
			xtick distance = 1,
			ytick distance = 1,
			restrict y to domain=-10:10,
			minor tick num = 1,
			major grid style = {lightgray},
			minor grid style = {lightgray!25},
			width = \linewidth,
			height = 0.7\linewidth,
			xlabel = {$\Delta V/U$},
			ylabel = {$\kappa_\mathrm{Mott}$,$\kappa_\mathrm{semi}$},
			legend cell align = {right},
			legend style={at={(.95,.3)},anchor=east},
			]

				\addplot+[
		thin, mark size=2pt, blue, dashed,mark repeat=5,mark phase=1, mark=*,mark options={scale=1, fill=blue}
			] table[x index = 0, y index = 1] {pics/vgl_decay_T02_v15_mott_semi.dat};

            \addplot+[
		thin, mark size=2pt,mark repeat=5,mark phase=1, mark=square*, red, mark options={scale=1, fill=red}
		] table[x index = 0, y index = 1,  skip coords between index={17}{1000}] {pics/vgl_decay_T02_v02_mott_semi.dat};
			
		\addplot+[
			thin, mark size=2pt, blue, dashed,mark repeat=5,mark phase=1, mark=triangle*, mark options={scale=1, fill=blue}
			] table[x index = 0, y index = 2] {pics/vgl_decay_T02_v15_mott_semi.dat};

		\addplot+[
		thin, mark size=2pt,mark repeat=5,mark phase=2, mark=diamond*, mark options={scale=1, fill=red}, red
		] table[x index = 0, y index = 2,  skip coords between index={17}{1000}] {pics/vgl_decay_T02_v02_mott_semi.dat};
			\legend{$\kappa_\mathrm{Mott}$ $V=1.5U$, $V=0.2U$, $\kappa_\mathrm{semi}$ $V=1.5U$, $V=0.2U$}

             \addplot+[
		thin, mark size=2pt,mark repeat=5,mark phase=1, mark=square*, red, mark options={scale=1, fill=red}
		] table[x index = 0, y index = 1,  skip coords between index={0}{17}] {pics/vgl_decay_T02_v02_mott_semi.dat};
			\addplot+[
		thin, mark size=2pt,mark repeat=5,mark phase=2, mark=diamond*, mark options={scale=1, fill=red}, red
		] table[x index = 0, y index = 2,  skip coords between index={0}{17}] {pics/vgl_decay_T02_v02_mott_semi.dat};
			
		\end{axis}
		
	\end{tikzpicture}
	\caption{Decay constants \red{$\kappa_\mathrm{Mott}$,$\kappa_\mathrm{semi}$} for both sides with $V=0.2U$ (red) and $V=1.5U$ (blue dashed) with $T=0.2U$ in the unpolarized case. Note the gap in the red square and diamond curves.}
	\label{fig:deconst_mott_semi_vgl}
\end{figure}

\begin{figure}
	\centering
	\begin{tikzpicture} 
		\begin{axis}[
			xmin = -4, xmax =0,
			ymin = 0, ymax = 6,
			xtick distance = 1,
			ytick distance = 1,
			restrict y to domain=-10:10,
			minor tick num = 1,
			major grid style = {lightgray},
			minor grid style = {lightgray!25},
			width = \linewidth,
			height = 0.7\linewidth,
			xlabel = {$\Delta V/U$},
			ylabel = {$\kappa_\mathrm{Mott}$, $\kappa_\mathrm{semi}$},
			legend cell align = {right},
			legend style={at={(.45,.3)},anchor=east}
			]

			\addplot+[
		thin, mark size=2pt,only marks, gray,mark options={scale=1, fill=gray}, mark=*,mark repeat=5,mark phase=1
		] table[x index = 0, y index = 1] {pics/decay_const_one_int_vm0p4_T0p2_negdV.dat};
		
			\addplot+[
		thin,red, mark size=2pt,mark options={scale=1, fill=red},only marks, mark=square*,mark repeat=5,mark phase=1
		] table[x index = 0, y index = 1] {pics/decay_const_one_int_v0p7_T0p2_negdV.dat};
		
			\addplot+[
		thin,blue, mark size=2pt,mark options={scale=1, fill=blue},only marks, mark=triangle*,mark repeat=5,mark phase=1
		] table[x index = 0, y index = 1] {pics/decay_const_one_int_v1p3_T0p2_negdV.dat};

					\addplot+[
		thin,gray, mark size=2pt,mark options={scale=1, fill=gray},only marks, mark=diamond*,mark repeat=5,mark phase=1
		] table[x index = 0, y index = 2] {pics/decay_const_one_int_vm0p4_T0p2_negdV.dat};
		
			\addplot+[
		thin,red, mark size=2pt,mark options={scale=1, fill=red},only marks, mark=pentagon*,mark repeat=5,mark phase=1
		] table[x index = 0, y index = 2] {pics/decay_const_one_int_v0p7_T0p2_negdV.dat};
		
			\addplot+[
		thin,blue, mark size=2pt,mark options={scale=1, fill=blue},only marks, mark=asterisk,mark repeat=5,mark phase=1
		] table[x index = 0, y index = 2] {pics/decay_const_one_int_v1p3_T0p2_negdV.dat};

			\legend{$\kappa_\mathrm{Mott}$ $V=-0.4U$,$0.7U$,$1.3U$, $\kappa_\mathrm{semi} $ $V=-0.4U$,$0.7U$,$1.3U$}

		\end{axis}
		
	\end{tikzpicture}
	\caption{Decay constants \red{$\kappa_\mathrm{Mott}$,$\kappa_\mathrm{semi}$} for a negative interface perturbation $\Delta V$ in the unpolarized case. The different colors depict different on-site potentials $V$. Hopping strength $T=0.2U$.}
	\label{fig:deconst_semi_v_negDV}
\end{figure}

In the case of a negative interface perturbation, essentially all the characteristics observed for positive interface perturbations reappear, see Fig.~\ref{fig:deconst_semi_v_negDV}.

Experimental measurements at the $\mathrm{LaTiO}_3/\mathrm{SrTiO}_3$ system show a tunnelling strength of $T=0.3\mathrm{eV}$ with a Coulomb repulsion of $U=6-20T$, such that realistic values are in the $T=0.16-0.05U$ 
regime. This matches the here used value of $T=0.2U$. \red{The decay constant for these parameters yields bound states with a non-zero quasi-particle density extending a few layers around the interface, consistent with other works \cite{Maznichenko2020,Maznichenko2019}.}
The bound states found here have dispersion in the parallel 
direction. It is likely that all these subbands are partially filled leading to a metallic state parallel to the interface as it was measured in \cite{ohtomo2002artificial,Maznichenko2024,okamoto2004electronic}.

\section{Unpolarized Mott region between two semiconductors}
Utilizing the same methodology, we can also compute bound states \red{for the quasi-particle wave functions} in a system with two interfaces, where an unpolarized Mott insulator is positioned between two semi-infinite semiconductors, similar to a quantum well. Notably, there is no pre-requisite for the semiconductors to be identical; they can have distinct \red{band offsets} \(V\) and interface perturbations \(\Delta V\). The Mott region now spans \red{the site indices} \(-a \leq \mu \leq a\), leading to the following \textit{ansatz}:

\begin{equation}
p_n^0= \begin{cases}
	A s_1^\mu & \mu<-a, \\
	B r_+^\mu + C r_-^\mu & -a \leq \mu \leq a ,\\
	D s_2^\mu & a<\mu.
\end{cases}
\end{equation}
Here, \(|s_1| >1\) and \( |s_2| <1\), with no requirement on the absolute value in the Mott region. \red{Remember that the $s_i$ give the electron wave function eigenmodes in the semiconductor while the $r_i$ give the quasi-particle eigenmodes in the Mott insulator.}
The four boundary conditions now yield two conditions relating the amplitudes:
\begin{equation}
\begin{aligned}
	\label{eq:amplitudes_twointer}
	1&=\frac{2(E-U)}{2E-U}\frac{s_1^{a+1}}{A} \left(B r_+^{-a-1}+C r_+^{a+1}\right), \\
	1&=\frac{2(E-U)}{2E-U}\frac{s_2^{-a-1}}{D} \left(B r_+^{a+1}+C r_+^{-a-1}\right). 
\end{aligned}
\end{equation}
Additionally, there are two defining equations for the interface perturbations at the interfaces:
\begin{equation}
\begin{aligned}
	\label{eq:con_pot_twointer}
	\frac{\Delta V_1 Z}{T}&=\left[ \frac{2(E-U)}{2E-U} \frac{B r_+^{-a}+C r_+^{a}}{A}-s_1^{-a}\right]s_1^{a+1}, \\
	\frac{\Delta V_2 Z}{T}&=\left[ \frac{2(E-U)}{2E-U} \frac{B r_+^{a}+C r_+^{-a}}{D}-s_2^{a}\right]s_2^{-a-1}.
\end{aligned}
\end{equation}
In the symmetric case, where the left and right semiconductors are the same, \(s_2=s_{\bar{1}}\) holds to satisfy the condition of absolute values (with the bar denoting the opposite index). \red{This ensures that the bound state does decay inside both semiconducting half-spaces.}

Now, there exist even (\(+\)) solutions (with \(B=C\), \(A=D\)) and odd (\(-\)) solutions (with \(B=-C\), \(A=-D\)). With this, the boundary conditions for the coefficients and the defining equation for bound states in this system are given by:

\begin{align}
	1&=\frac{2(E-U)}{2E-U}\frac{B}{A} \left( r_+^{-a-1}\pm r_+^{a+1}\right)s_1^{a+1}, \\
	\frac{\Delta V Z}{T}+s_1&= \frac{r_+^a \pm r_+^{-a}}{r_+^{a+1}\pm r_+^{-a-1}}.
\end{align}
Here, \(+\) denotes the even parity case, while \(-\) indicates the odd one. \red{The energy of the bound state physically fixes the mathematical \textit{ansatz} for the $s_i$. \(s_1= s_+\) yields energies greater than the semiconducting band, while \(s_-\) yields those below. The second equation is the defining one for quasi-particle bound states in this double interface system, the first one normalizes the wave function amplitudes.} 


\subsection{Standing Wave Solutions}
\begin{figure}
   	\begin{tikzpicture}
	\begin{axis}[
		xmin = -0.5, xmax = 1.5,
		ymin = -4, ymax = 10,
restrict y to domain=-4:20,
		xtick distance = .4,
		ytick distance = 3,
		minor tick num = 1,
		major grid style = {lightgray},
		minor grid style = {lightgray!25},
		width = \linewidth,
		height = 0.7\linewidth,
		ylabel = {L.S and R.S of Eq.(27) },
		xlabel = {$E/U$},
		legend cell align = {right},
        title={graphical solution}
		]

		\addplot[
		smooth,
		thin, line width=1.5pt
		] table[x index=0, y index=2] {pics/def_eq_evenoddparitysol.dat};
		
			\addplot[
		smooth,
		thin,
		dashed, red, line width=1.5pt
		] table[x index=0, y index=3] {pics/def_eq_evenoddparitysol.dat};
		
			\addplot[
		smooth, line width=1.5pt,
		dotted, blue
		] file {pics/def_eq_evenoddparitysol.dat};
		
		\legend{even R.S, odd R.S, L.S}
		
	\end{axis}
\end{tikzpicture}
\caption{Defining equation (27) with $\Delta V=-0.6U$, $T=0.4U$ and $V=1.5U$ in the zero parallel momentum sector and $a=6$. The crossings of the solid and dashed line with the dotted one give the energies for the even and odd parity solutions, respectively.}
\label{fig:two_inter_defeq}
\end{figure}

Disregarding any interface perturbation, the \red{bound state }energy 
must fall into one of the two Mott bands (see Fig.~\ref{fig:two_inter_defeq}).
Defining the Mott region as $\mu \in [-a, a]$, i.e. with $2a+1$ Mott lattice sites, \red{there are even and odd parity solutions for the quasi-particle wave function. This is well known from the standard quantum well.} 
\red{Mathematically, }we obtain $2a+2$ even-parity solutions \red{for the wave function}, which are evenly distributed into $a+1$ solutions in both the lower and upper Hubbard bands. For odd parity, there are $2a$ solutions, equally divided into $a$ solutions in the lower and upper Hubbard bands. 
All of them have quasi-particle density leaking from the Mott into the semiconductors.

\red{The physical condition of the wave function decay away from the interfaces as well as the band offset $V$ set the mathematical \textit{ansatz} for $s_1$.}
A change in the \red{offset} has a minimal impact on the solutions, slightly lowering the energy (analogous to a potential well), as well as on the leaking. For solutions in the lower Hubbard band, the leakage is nearly independent of the \red{band offset}.

The \red{doublon and holon wave functions} manifest as standing waves with $|e^{\pm i \kappa_\mathrm{Mott}\mu}|=1$ in the Mott region. \red{As known from the quantum well, even and odd parity solutions alternate in energy.} \red{Hence,} in both the lower and upper Hubbard bands, the \red{standing waves} have energies \(E_0^L, E_1^L, \ldots, E_{2a}^L\), \red{and \(E_0^U, E_1^U,\ldots E_{2a}^U\) respectively}, where even (odd) \red{index energies} correspond to even (odd) parity solutions. 
\red{Additionally, and again analogous to the quantum well, the }number of \red{nodes} of \red{the holon wave function} \(p_\mu^0\) increases with energy from 0 to \(2a\).

\red{More precisely, }the corresponding quasi-particle probability distributions \(|p_\mu^0(E_n)|^2\) have \(2(\frac{a}{2}-|\frac{a}{2}-\frac{n}{2}|)\) nodes for $n$ even. For \(a=6\), this results in the sequence \(0-2-4-6-4-2-0\). Performing the same analysis for the odd \red{index} energy solutions
yields \(2(\frac{a}{2}-|\frac{a}{2}-\frac{n}{2}|)+1\) nodes, (\(1-3-5-5-3-1\) for \(a=6\)). In the lower Hubbard band the quasi-particle probability distribution is \red{the same} between states with the same number of nodes \red{independent of their energy}. In the upper Hubbard band, \red{this is not true anymore}. As the energy of the solution increases, more of the probability becomes localized at the interfaces; thereby the amount of leaked density increases as well.

Bound state solutions persist within this heterostructure for all \(V\), \red{independent of the band offset}.
As in the single-interface case, the bound states display a dispersion in the parallel direction, leading to subbands. These subbands are partially filled and, as there is quasi particle weight at the interface, the bound states exhibit metallic behavior there \cite{ohtomo2002artificial,okamoto2004electronic,Maznichenko2024}.

\begin{figure}
	\begin{tikzpicture}
		\begin{axis}[
			xmin = -10, xmax = 10,
			ymin = -.2, ymax = .3,
			xtick distance = 5,
			ytick distance = .2,
			minor tick num = 1,
			major grid style = {lightgray},
			minor grid style = {lightgray!25},
			width = \linewidth,
			height = 0.7\linewidth,
			ylabel = {holon wave function $p_\mu^0$},
			xlabel = {site index $\mu$},
			legend cell align = {right},
			legend columns=-1
			]

			\addplot[
			smooth, line width=2pt
			] file {pics/two_inter_nodV_solutions_lowerband_E0_E3_T0.4_pp0_U1_V15_dV0_Z4_a6.dat};

			\addplot[
			smooth,
			dashed, red   , line width=2pt
			] table[x index=0, y index=2] {pics/two_inter_nodV_solutions_lowerband_E0_E3_T0.4_pp0_U1_V15_dV0_Z4_a6.dat};
			
				\addplot[
			smooth,
			dotted, blue, line width=2pt
			] file {pics/two_inter_nodV_solutions_upperband_E2_E7_T0.4_pp0_U1_V15_dV0_Z4_a6.dat};

			\addplot[
			smooth,
			dashdotted, orange , line width=2pt
			] table[x index=0, y index=2] {pics/two_inter_nodV_solutions_upperband_E2_E7_T0.4_pp0_U1_V15_dV0_Z4_a6.dat};
			
			\draw (-6,-1)-- (-6,1);
						\draw (6,-1)-- (6,1);

									\node[draw] at (-8,-.15) {Semi};
																		\node[draw] at (8,-.15) {Semi};
																											\node[draw] at (0,-.15) {Mott};
									
			\legend{$E_{0}^L$, $E_{4}^L$,$E_{2}^U$,$E_{12}^U$}
			
		\end{axis}
	\end{tikzpicture}
	\caption{Bound states in the lower and upper Hubbard band with $\Delta V=0$, $T=0.4U$ and $V=1.5U$ in the zero parallel momentum sector and $a=6$.}
	\label{fig:no_dv_twointer_example}
\end{figure}

\subsection{With Interface Perturbation}
Introducing an interface perturbation results in one (or in special cases two) additional bound state(s) 
with \(e^{i \kappa_\mathrm{Mott}\mu} \neq 1\). \red{In the Mott region these are not standing waves, but the wave function amplitude decays away from the interface, see Fig.~\ref{fig:add_bound_statesdV_evenp}.}
The interface perturbation must exceed a minimum value for these additional bound states to exist, which is given by:
\begin{equation}
\Delta V_\mathrm{min}=\frac{T}{Z}\left( -1+\frac{\cos(a \kappa_\mathrm{Mott}(E_{SL}))}{\cos((a+1) \kappa_\mathrm{Mott}(E_{SL}))}  \right)
\end{equation}
with \(E_{SL}= \pm 2T/Z - T_\mathbf{k}^\| +V \). \red{The $\pm$ moves the bound state below ($-$) or above ($+$) the semiconducting band.}
These additional bound states are illustrated in Fig.~\ref{fig:add_bound_statesdV_evenp}.  

We note that two (rather than one) extra states emerge 
directly below and above one of the Hubbard bands 
if \red{the band offset }$V$ and \red{the perturbation }$\Delta V$ align favorably. 
The lower state originates from the lowest solution initially inside the band being pushed outward.

The increase in the magnitude of the interface perturbation enhances localization at the interface, resulting in an increased decay constant in the semiconductors. The standing wave solutions within the Mott bands exhibit only mild sensitivity to the interface perturbation.

\begin{figure}
	\begin{tikzpicture}
		\begin{axis}[
			xmin = -10, xmax = 10,
			ymin = -.2, ymax = .3,
			xtick distance = 5,
			ytick distance = .2,
			minor tick num = 1,
			major grid style = {lightgray},
			minor grid style = {lightgray!25},
			width = \linewidth,
			height = 0.7\linewidth,
			ylabel = {holon wave function $p_\mu^0$},
			xlabel = {site index $\mu$},
			legend cell align = {right}
			]

			\addplot[
			smooth,
			thin, line width=2pt
			] file {pics/two_inter_withdV_T0.4_pp0_U1_V1.5_dV-0.25_one_-0.35_two_pushedout_Z4_a6.dat};

			\addplot[
			smooth,
			thin,
			dashed, red   , line width=2pt
			] table[x index=0, y index=2]  {pics/two_inter_withdV_T0.4_pp0_U1_V1.5_dV-0.25_one_-0.35_two_pushedout_Z4_a6.dat};
			
			\addplot[
			smooth,
			thin,
			dotted, blue, line width=2pt
			] table[x index=0, y index=3]  {pics/two_inter_withdV_T0.4_pp0_U1_V1.5_dV-0.25_one_-0.35_two_pushedout_Z4_a6.dat};
			
				\draw (-6,-1)-- (-6,1);
			\draw (6,-1)-- (6,1);

			\node[draw] at (-8,-.15) {Semi};
			\node[draw] at (8,-.15) {Semi};
			\node[draw] at (0,-.15) {Mott};

			\legend{$\Delta V=-0.25U$,$\Delta V=-0.35U$ lower,$\Delta V=-0.35U$ upper}
			
		\end{axis}
	\end{tikzpicture}
	\caption{Additional bound state solutions for different interface perturbations $\Delta V=-0.25U$ and $-0.35U$. The latter one has two bound states, as the lowest standing wave solutions is pushed out of the band. $T=0.4U$ and $V=1.5U$ in the zero parallel momentum sector and $a=6$.}
	\label{fig:add_bound_statesdV_evenp}
\end{figure}

\subsection{Non-Symmetric Case}
\red{The band offset of the semiconductors coupled to the Mott insulator do not necessarily need to be the same.} In the case of asymmetry, where \(V_1 \neq V_2\) and \(\Delta V_1 \neq \Delta V_2\), three out of the four Eq.~\ref{eq:amplitudes_twointer}, Eq.~\ref{eq:con_pot_twointer} govern the amplitudes \(A\), \(B\), \(C\), and \(D\), with one serving as the normalization constant. The remaining equation becomes the determining equation for the energy of the bound state solutions.
These solutions are also asymmetric. In the absence of an interface perturbation, the half-space with the lower \red{band offset} accumulates slightly more quasi-particle probability.

\section{Single Interface at Mott-N\'eel Background}
\red{As discussed in Sec.~\ref{subsec:unpolMott} and Sec.~\ref{subsec:bipartMott}, 
one may consider two mean-field background solutions in the Mott insulator. Previously we discussed the unpolarized Mott state without any spin ordering
whereby the doublons and holons may simultaneously live on the same lattice site. 
This is in contrast to the bi-partite Mott-N\'eel state 
where the doublon and holon wave functions are 
exclusively non-zero on one of the sublattices, but 
remain coupled across lattice sites. In the setting chosen here, the holon wave function only lives on sublattice $A$ ($p_\mu^{0A}\neq 0$ in Mott), while the doublon one lives on sublattice $B$ ($p_\mu^{1B} \neq 0$ in Mott).} In this antiferromagnetic Mott-N\'eel background, the interface bound states generally comprise both the even and odd solution. Choosing $\mu \leq 0$ as the semiconducting and $\mu>0$ as the Mott insulating sites, the \textit{ansatz} \red{for the quasi-particle wave function} is 
\begin{equation}
\label{eq:MNAnsatzSemi}
		\psi_{\mu \leq 0}=\begin{bmatrix}
		p_\mu^{0A} \\
		p_\mu^{1A} \\
		p_\mu^{0B} \\
		p_\mu^{1B} \\
	\end{bmatrix} = \frac{1}{2} A \lambda_+^n \begin{bmatrix}
		1 \\ 0\\1\\0
	\end{bmatrix}+\frac{1}{2}B \rho_-^n \begin{bmatrix}
		1 \\ 0\\-1\\0 \end{bmatrix}
\end{equation}	
	\red{in the semiconducing sites. It is chosen to represent the conduction band.}
    \red{In the Mott insulating half-space it reads:}
	\begin{equation}
    \label{eq:MNAnsatzMott}
		\psi_{\mu>0}=  \frac{1}{\sqrt{2}} \kappa_i^n \begin{bmatrix} A_i \\ 0 \\0 \\ B_i \end{bmatrix} + \frac{1}{\sqrt{2}} \kappa_j^n \begin{bmatrix} A_j \\ 0 \\0 \\ B_j \end{bmatrix}.
	\end{equation}
    Inserting this \textit{ansatz} into the boundary conditions Eq.~\ref{eq:boundarycondMN} yields:
	\begin{equation}
		\begin{aligned}
			-\frac{Z}{T}\Delta V \frac{1}{2}(A-B)+A \lambda_++B\rho_-&=\frac{1}{\sqrt{2}}B_i \kappa_i+\frac{1}{\sqrt{2}}B_j \kappa_j \\
			A-B=&\frac{1}{\sqrt{2}}B_i+\frac{1}{\sqrt{2}}B_j \\
			-\frac{Z}{T}\Delta V \frac{1}{2}(A+B)+A \lambda_+-B\rho_-&=\frac{1}{\sqrt{2}}A_i \kappa_i+\frac{1}{\sqrt{2}}A_j \kappa_j \\
			A+B=&\frac{1}{\sqrt{2}}A_i+\frac{1}{\sqrt{2}}A_j \\
		\end{aligned}
	\end{equation}
These equations relate the interface perturbation $\Delta V$ to the amplitudes
\red{$A$ and $B$.} 
As in the unpolarized case, there is a minimally needed strength 
of the additional interface perturbation for bound state to exist, but, in contrast, this  threshold 
is constant at $\Delta V_\mathrm{min} \approx \frac{2T}{Z}$. Another distinct feature is that, for bound states with an energy below the semiconducting band, only a positive interface perturbation $\Delta V >0$ supports bound states. Even though the charge background with one electron per site is the same, the different spin backgrounds lead to distinctive behavior of the bound states.

\section{Conclusions}
Utilizing the hierarchy of correlations alongside the Fermi-Hubbard model we derived equations for the existence of bound states \red{of the doublon and holon wave function} at single and multiple interface systems with different spin backgrounds. We account for an additional interface perturbation.
These quasi-particle bound states manifest in the semiconductor as real electrons, while in the Mott insulator they are doublons and holons. 
Even though we used a model Hamiltonian, realistic values for the on-site potential,
 the hopping strength and the Coulomb repulsion can be extracted from literature.

At a solitary interface, the spin background 
and the charge backgrounds 
cannot be discussed independently: Both in the unpolarized case with (on average) one electron per site and  in the antiferromagnetic Mott-N\'eel background an interface perturbation is needed to support bound states. The threshold of the perturbation strength depends on the spin background;  in the unpolarized case it additionally  depends on the band alignment, whereas in the Mott-N\'eel case it is constant.

In the case of a double-interface system, bound states exist even without an interface perturbation and manifest as standing wave solutions. 
The introduction of the interface perturbation adds additional states that are highly localized at the interface. 

For future investigations, exploring second-order effects would be valuable. In the next hierarchical order of correlations, spin-spin and doublon-holon correlations emerge. They effectively renormalize the quasi particle energies used here, but they also potentially might have their own bound state structure.
Additionally, studying the back reaction of correlation functions on the mean-field background could unveil insights, including potential space charge layer formation and effects of electron-density variation by spreading across the interface.

\section*{Acknowledgments}
The authors thank F. Queisser and R. Schützhold for fruitful discussions and valuable feedback on the manuscript. 
This work is funded by the Deutsche Forschungsgemeinschaft (DFG, German Research Foundation) – Project-ID 278162697– SFB 1242.
\appendix

\section{Hierarchy of Correlations}
\label{app:hierarchyofcorrelations}
In order to describe the (quasi-)particles in the heterostructure, we resort to the hierarchy of correlations  \cite{Queisser2014,Queisser2019,PhysRevA.82.063603}. 
For any lattice Hamiltonian of the form:
\begin{equation}
	\label{eq:generalManyBodyHam}
	\h H = \frac{1}{Z} \sum_{\mu \nu}\h H_{\mu \nu} + \sum_\mu \h H_\mu
\end{equation}
we can find an infinite set of equations for the density operator $\h \rho$. In this, $\mu$ and $\nu$ are generalized coordinates. In the regime of large coordination number $Z \gg 1$ a truncation scheme is applicable to give a closed set of equations and an iterative way to solve this. 

The starting point is the Heisenberg equation: 
\begin{equation}i \partial_t \h \rho = \big[H,\h \rho \big]= \frac{1}{Z}\sum_{\mu \nu} \widehat{\mathcal{L}}_{\mu\nu}\h \rho + \sum_\mu \widehat{\mathcal{L}}_\mu \h \rho
\end{equation}
with the Liouville superoperators $\widehat{\mathcal{L}}_{\mu\nu}=\left[\h H_{\mu\nu},\h \rho\right]$ and $\widehat{\mathcal{L}}_\mu = \comm{\h H_\mu}{\h \rho}$. The next step is the decomposition of the density operator. 
Any operator whose expectation value one might be interested in is computed from a subset of lattice sites. This allows for the decomposition into on-site and correlated parts:
\begin{equation}
	\begin{aligned}
		\h \rho_{\mu\nu} &= \h \rho^\text{corr}_{\mu\nu}+ \h \rho_\mu \h \rho_\nu, \\
		\h \rho_{\mu\nu\lambda}&=\h \rho^\text{corr}_{\mu\nu\lambda}+\h \rho^\text{corr}_{\mu\nu}\h \rho_\lambda +\h \rho^\text{corr}_{\mu\lambda}\h \rho_\nu +\h \rho^\text{corr}_{\nu\lambda}\h \rho_\mu + \h \rho_\mu \h\rho_\nu\h\rho_\lambda,
	\end{aligned}
\label{eq:splittingcorr}
\end{equation}
and so on. In order to get the time evolution of these, we need to calculate:
\begin{equation}
    i \partial_t \h\rho_\mu =\frac{1}{Z}\sum_{\alpha \neq\mu} \mathrm{tr}_\alpha\left(\widehat{\mathcal{L}}^S_{\alpha\mu} \left[ \h \rho_{\mu\alpha}^\text{corr}+\h \rho_\alpha \h\rho_\mu\right]\right) + \widehat{\mathcal{L}}_\mu \h\rho_\mu.
\end{equation}
with the symmetrized from $\widehat{\mathcal{L}}^S_{\mu\nu}=\widehat{\mathcal{L}}_{\mu\nu}+\widehat{\mathcal{L}}_{\nu\mu}$ for the on-site density operator. This time evolution contains the two-point correlator $\h \rho_{\mu \nu}^\mathrm{corr}$.
The same is done for the two-site density operator, such that we can combine these two results to find:
\begin{equation}
	\begin{aligned}
		i \partial_t \h\rho^\text{corr}_{\mu\nu}=& \widehat{\mathcal{L}}_\mu \h\rho^\text{corr}_{\mu\nu}+\frac{1}{Z}\widehat{\mathcal{L}}_{\mu\nu}\left(\h\rho^\text{corr}_{\mu\nu}+\h\rho_\mu\h\rho_\nu\right) 
		\\&-\frac{\h\rho_\mu}{Z}\tr_\mu\left(\widehat{\mathcal{L}}^S_{\mu\nu}\left[\h\rho^\text{corr}_{\mu\nu}+\h\rho_\mu\h\rho_\nu\right]\right)\\
		&+\frac{1}{Z}\sum_{\alpha \neq\mu\nu}\tr_\alpha\left(\widehat{\mathcal{L}}_{\mu\alpha}^S\left[\h\rho^\text{corr}_{\mu\nu\alpha}+\h\rho^\text{corr}_{\mu\nu}\h\rho_\alpha+\h\rho^\text{corr}_{\nu\alpha}\h\rho_\mu\right]\right)
		\\&+ (\mu \leftrightarrow \nu)
	\end{aligned}
	\label{time_evo_rho_munucorr}
\end{equation}
as the time evolution of the two-point correlator $\h \rho_{\mu \nu}^\mathrm{corr}$. It contains the three-point correlator $\h \rho_{\mu \nu \lambda}^\mathrm{corr}$. This builds up a set of equations:
\begin{eqnarray}
	\label{evolution_appendix}
	i\partial_t \hat\rho_\mu 
	&=& 
	F_1(\hat\rho_\mu,\hat\rho_{\mu\nu}^{\mathrm{corr} })
	\,,\nonumber\\
	i\partial_t \hat\rho_{\mu\nu}^{\mathrm{corr} } 
	&=& 
	F_2(\hat\rho_\mu,\hat\rho_{\mu\nu}^{\mathrm{corr} },\hat\rho_{\mu\nu\lambda}^{\mathrm{corr} })
	\,,
	\nonumber\\
	i\partial_t \hat\rho_{\mu\nu\lambda}^{\mathrm{corr} } 
	&=& 
	F_3(\hat\rho_\mu,\hat\rho_{\mu\nu}^{\mathrm{corr} },\hat\rho_{\mu\nu\lambda}^{\mathrm{corr}},
	\hat\rho_{\mu\nu\lambda\kappa}^{\mathrm{corr} })
		\,,
	\nonumber\\
	i\partial_t \hat\rho_{\mu\nu\lambda \alpha}^{\mathrm{corr} } 
	&=& 
	F_4(\hat\rho_\mu,\hat\rho_{\mu\nu}^{\mathrm{corr} },\hat\rho_{\mu\nu\lambda}^{\mathrm{corr}},
	\hat\rho_{\mu\nu\lambda\kappa}^{\mathrm{corr} },	\hat\rho_{\mu\nu\lambda\kappa\beta}^{\mathrm{corr} }).
\end{eqnarray}
The specific form of the functionals $F_n$ is dictated by the Hamiltonian.

 If the initial state satisfies scaling relations such that \(\ell$-point correlations are of order \(\mathcal{O}(Z^{-\ell+1})\), this scaling persists for all times \cite{queisser2023hierarchy,Queisser2014}. Exploiting the scaling behavior, we approximate the equations to zeroth and first order, yielding:
\begin{equation}
\begin{aligned}
	i\partial_t \hat{\rho}_\mu &\approx F_1(\hat{\rho}_\mu, 0), \quad \text{with solution} \quad \hat{\rho}_\mu^0, \\
	i\partial_t \hat{\rho}_{\mu\nu}^{\text{corr}} &\approx F_2(\hat{\rho}_\mu^0, \hat{\rho}_{\mu\nu}^{\text{corr}}, 0).
\end{aligned}
\end{equation}
These two equations are used to describe the charge modes within the studied system.
$\hat{\rho}_\mu^0$ encodes the mean-field background charge and spin structure.
\section{Hierarchy for the Fermi Hubbard Model}
\label{app:hierarchyforFermiHubbard}
In order to apply the hierarchy of correlations to the Fermi-Hubbard model Eq.~\ref{eq:FHMHamiltonian},
we first introduce the two different spin backgrounds. There is the unpolarized state:
\begin{equation}
\h \rho_\mu^0=\frac{\ket{\uparrow}_\mu\!\bra{\uparrow}+\ket{\downarrow}_\mu\!\bra{\downarrow}}{2}    
\end{equation}
and there is the antiferromagnetic Mott-N\'eel state with its two sublattices $A$ and $B$ arranged in a checkerboard structure:
\begin{equation}
	\h \rho_\mu^0 = \begin{dcases}
\ket{\uparrow}_\mu\!\bra{\uparrow} & \mu \in A, \\
\ket{\downarrow}_\mu\!\bra{\downarrow} & \mu \in B.
	\end{dcases}
\end{equation}
Independent of the mean-field background, it is instructive to introduce quasi-particle operators, similar to the idea of Hubbard X \cite{hubbard1965electron,ovchinnikov2004hubbard} or composite operators \cite{mancini2004hubbard}, as:
\bea
\hat c_{\mu s I}=\hat c_{\mu s}\hat n_{\mu\bar s}^I=
\left\{
\begin{array}{ccc}
	\hat c_{\mu s}(1-\hat n_{\mu\bar s}) & {\rm for} & I=0, 
	\\ 
	\hat c_{\mu s}\hat n_{\mu\bar s} & {\rm for} & I=1
\end{array}
\right.
\ea
for doublons \(I=1\) and holes \(I=0\). These better describe the physics, but note that these operators are approximately, but not exactly equal to the quasi-particle creation and annihilation operators for holons and doublons, see, e.g. \cite{Avigo2020}. The label $\bar{s}$ denoted the spin index opposite to $s$. 
For the correlation functions $\langle \h c_{\mu s I}\h c_{\nu s J} \rangle$ we find:
\begin{equation}
\begin{aligned}
\label{corr-evolution}
i\partial_t
\langle\hat c^\dagger_{\mu s I}\hat c_{\nu s J}\rangle^{\mathrm{corr}}
=
\frac1Z\sum_{\lambda L} T_{\mu\lambda}
\langle\hat n_{\mu\bar s}^I\rangle^0
\langle\hat c^\dagger_{\lambda s L}\hat c_{\nu s J}\rangle^{\mathrm{corr}}
\\
-
\frac1Z\sum_{\lambda L} T_{\nu\lambda}
\langle\hat n_{\nu\bar s}^J\rangle^0
\langle\hat c^\dagger_{\mu s I}\hat c_{\lambda s L}\rangle^{\mathrm{corr}}
\\
+
\left(U_\nu^J-U_\mu^I+V_\nu-V_\mu\right) 
\langle\hat c^\dagger_{\mu s I}\hat c_{\nu s J}\rangle^{\mathrm{corr}}
\\
+\frac{T_{\mu\nu}}{Z}
\left(
\langle\hat n_{\mu\bar s}^I\rangle^0
\langle\hat n_{\nu s}^1\hat n_{\nu\bar s}^J\rangle^0
-
\langle\hat n_{\nu\bar s}^J\rangle^0
\langle\hat n_{\mu s}^1\hat n_{\mu\bar s}^I\rangle^0
\right) +\mathcal{O}(1/Z^2)
\end{aligned}
\end{equation}
Here, we used the abbreviation $U_\mu^I=IU_\mu$, i.e., $U_\mu^I=U_\mu$ for 
$I=1$ and $U_\mu^I=0$ for $I=0$.

The relevant part of this time evolution, which describes the dynamics, can further be simplified by a factorization \cite{navez2014quasi,Queisser2014}.
This yields the doublon ($I,J=1$) and holon ($I,J=0$) amplitudes: 
\begin{equation}
\langle \hat{c}_{\mu s I} \hat{c}_{\nu s J}\rangle^{\text{corr}} = (p_\mu^I)^* p_\nu^J \, .
\end{equation} 
The two amplitudes can be grouped together using a spinor notation and 
governing equations for these (quasi) particles \cite{verlage2024quasi}
can be derived. Assuming a highly symmetric lattice such as a hyper-cubic one allows us to perform the corresponding Fourier transform parallel to the interface by:
\begin{equation}
	\begin{aligned}
		p_{\mu s I}&=\frac{1}{\sqrt{N^\parallel}}\sum_{\mathbf{k}^\parallel}p_{n,\mathbf{k}^\parallel,s}^Ie^{i  \mathbf{k}^\parallel \cdot\mathbf{x}_\mu^\parallel},\\
		T_{\mu\nu}&=\frac{Z}{N^\parallel}\sum_{\mathbf{k}^\parallel} T_{m,n,\mathbf{k}^\parallel}e^{i \mathbf{k}^\parallel \cdot\left(\mathbf{x}_\mu^\parallel-\mathbf{x}_\nu^\parallel\right)}
	\end{aligned}
\end{equation}
For the isotropic nearest neighbor hopping $T^\|_n=T^\perp_{n,n-1}=T$ the components read:
\begin{equation}
\begin{aligned}
	T_{m,n,\mathbf{k}^\parallel}&=\frac{T^\parallel_{\mathbf{k}^\parallel}}{Z}\delta_{m,n}+
	\frac{T}{Z}
	(\delta_{n,n-1}+\delta_{n,n+1}),\\
	T_{\mathbf{k}^\parallel}^\parallel&=2 T \sum_{x_i}\cos(p_{x_i}^\parallel) \equiv Z T_\mathbf{k}^{\|}\,,
\end{aligned}	
\end{equation}
with the hopping contribution $T_{\bf k}^\|$.
With this, the doublon and holon amplitudes follow the coupled equations:
\begin{equation}
	\label{difference2}
\begin{aligned}
	\left(E-U_\mu^I-V_\mu\right)p_\mu^I + \langle\hat{n}_{\mu}^I\rangle^0\sum_J T_{\bf k}^\| p_\mu^J \\
	= -T\frac{\langle\hat{n}_{\mu}^I\rangle^0}{Z}\sum_J \left(p_{\mu-1}^J+p_{\mu+1}^J\right).
\end{aligned}
\end{equation}
The different possibilities of the mean-field background contribute by the expectation values \(\langle\hat{n}_{\mu}^I\rangle^0 = \text{Tr}(\hat{n}_\mu^I \hat{\rho}_\mu^0)\). In the unpolarized state we have $\langle\hat n_{\mu}^I\rangle^0 =1/2$, while in the antiferromagnetic Mott-N\'eel state we need another index for the $A$- and $B$-sublattice as one of these has $\langle \h  n_{\mu_X \uparrow}\rangle=0 $ and $\langle \h n_{\mu_X \downarrow}\rangle=1$, while the other one has them reversed $\langle \h n_{\mu_Y \uparrow}\rangle=1$, $\langle \h n_{\mu_Y \downarrow}\rangle=0$.

\section{Single Unpolarized Interface}
\label{App:single_interface}
At the single interface between the Mott insulator $\mu <0$ and the semiconductor $\mu \geq 0$ we can write down two boundary conditions from Eq.~\ref{difference2} with $\langle\hat n_{\mu <0}^I\rangle^0 =1/2$, $\langle\hat n_{\mu \geq 0}^1\rangle^0 =0$ and $\langle\hat n_{\mu \geq 0}^0\rangle^0 =1$. Together with the relation between particles and holes $p_{\mu< 0}^0 E=p_{\mu< 0}^1 (E-U)$ in the Mott, the boundary conditions read:
\begin{equation}
\begin{aligned}
\label{eq:app_boundaryconditions}
\left(E-V-\Delta V+T_{\mathbf{k}}^\|\right)p_0^0 =& -\frac{T}{Z}\left( p_1^0+\frac{2E-U}{2(E-U)}p_{-1}^0\right), \\
\left(E+\frac{2E-U}{2(E-U)} T_\mathbf{k}^\|\right)p_{-1}^0 =& -\frac{T}{Z} \left(p_0^0+\frac{2E-U}{2(E-U)}p_{-2}^0 \right).
\end{aligned}
\end{equation}
Additionally, two identities hold within each individual region:
\begin{equation}
\label{eq:identities}
\begin{aligned}
	E+T_\mathbf{k}^{\|} \frac{2 E-U}{2(E-U)} &= -\frac{T}{Z}\left(r_{a}+\frac{1}{r_a}\right) \frac{2 E-U}{2(E-U)}, \\
	\left(E-V+T_\mathbf{k}^{\|}\right) &= -\frac{T}{Z}\left(s_{b}+\frac{1}{s_{b}}\right).
\end{aligned}
\end{equation}
By combining the \textit{ansatz} Eq.~\ref{eq:ansatz}, the boundary conditions Eq.~\ref{eq:app_boundaryconditions}, and using the identities Eq.~\ref{eq:identities}, we derive a relation between the amplitudes:
\begin{equation}
\frac{2E-U}{2(E-U)}\frac{A_a}{B_b} = 1.
\end{equation}
Furthermore, we obtain an equation relating the interface perturbation to the amplitudes and half-space solutions \(r_a\) and \(s_b\):
\begin{equation}
\frac{\Delta V Z}{T} = \frac{2E-U}{2(E-U)}\frac{A_a}{B_b} \frac{1}{r_{a}} - \frac{1}{s_{b}}.
\end{equation}
These two equations combine to the defining equation for bound states at the unpolarized single interface Eq.~\ref{eq:delta_v_def}. 

\subsection{Defining Equation}
\label{App:single_interface_defEq}
The defining equation from these two then reads:
\begin{equation}
	0=\frac{1}{r_a}-\frac{1}{s_b}-	\frac{\Delta V Z}{T}.
\end{equation}
Without an interface perturbation, $\Delta V \equiv 0$, this equation is solved by $b \kappa_\mathrm{semi}=a \kappa_\mathrm{Mott} + 2\pi c$ with an integer $c \in \mathbb{Z}$. But, this would also result in 
$|e^{ai  \kappa_\mathrm{Mott}}|=|e^{bi  \kappa_\mathrm{semi}}|$  which is in conflict with the \textit{ansatz} Eq.~\ref{eq:ansatz}. So, there are no solutions without the interface perturbation.
Inserting $r_a$ and $s_b$ yields a function
\bea
\label{eq:fE}
f(E)&=&\frac{1}{r_a}-\frac{1}{s_b}-\frac{\Delta V Z}{T} \nn
&=&a i \sqrt{1-\frac{Z^2 \left(2 E \left(\frac{E}{U-2 E}+1\right)+T_\mathbf{k}^\|\right)^2}{4 T^2}} +\nn
&&+b i \sqrt{1-\frac{Z^2 (E+T_\mathbf{k}^\|-V)^2}{4 T^2}} \nn
&&+\frac{Z (E (U-2 V)+U V)}{4 E T-2 T U}-\frac{\Delta V Z}{T}
\eea
whose zeros give the bound states. $a$ and $b$ are the $+$ or $-$, respectively.
The imaginary \enquote{bells} coincide with the Mott and semiconductor band, respectively, and are not affected by the interface perturbation.
The term $-\Delta V Z/T$ shifts the real part of the defining equation up and down.
In Fig.~\ref{fig:why_dV} the defining equation for $r_a=r_+$ and $s_b=s_-$ is shown without and with a interface perturbation together with the absolute value $|r_a|$ and $|s_b|$. Without any interface perturbation the zeros of the real part are inside the non-zero imaginary part. So, by adding the interface perturbation it follows directly that any solution is outside of the bands and  has imaginary $\kappa_\mathrm{Mott}$ and $\kappa_\mathrm{semi}$. The conditions $|r_a|>1$ and $|s_b|<1$ then dictate on which side of the \enquote{bells} the solution needs to be, because the absolute value (at least for the semiconductor) is a monotone function of energy $E$ interrupted by the plateau of the band. For the Mott, this is also true, but at $E=U/2$ there the absolute value goes down to zero again. 

\begin{figure}
	\centering
	\begin{tikzpicture} 
		\begin{axis}[
			xmin = .5, xmax =1.5,
			ymin = -2, ymax = 10,
			xtick distance = .25,
			ytick distance = 2,
			restrict y to domain=-10:10,
			minor tick num = 1,
			major grid style = {lightgray},
			minor grid style = {lightgray!25},
			width = \linewidth,
			height = 0.7\linewidth,
			xlabel = {$E/U$},
			ylabel = {$f(E)$},
			legend cell align = {right},
			cycle list name=black white,
			legend style={at={(.7,.8)},anchor=east},
			legend columns=2,
			grid=both
			]

			\addplot[
			] table[x index = 0, y index = 1] {pics/why_dV2.dat};

			\addplot[
			dashed
			]  table[x index = 0, y index = 2] {pics/why_dV2.dat};
			
			\addplot[
			dotted
			] table[x index = 0, y index = 3] {pics/why_dV2.dat};
			
			\addplot[
			dashdotdotted
			] table[x index = 0, y index = 4] {pics/why_dV2.dat};

			\addplot[
			densely dashed
			] table[x index = 0, y index = 5] {pics/why_dV2.dat};

			\legend{$R_0$, $I$, $R_1$, $|r_+|$,$|s_-|$}

		\end{axis}
		
	\end{tikzpicture}
	\caption{ Eq.~\ref{eq:fE} with $r_a=r_+$ and $s_b=s_-$. The solid line gives the real part without any interface perturbation $R_0=\mathrm{Re}f(E,\Delta V=0)$, the dashed line the imaginary part $I=\mathrm{Im} f(E)$ and the dotted the real part with non-zero interface perturbation $R_1=\mathrm{Re}f(E,\Delta V=0.2U)$. The parameters are $V=1.1U$, $T=0.4U$}
	\label{fig:why_dV}
\end{figure}

\subsection{Minimum Interface Perturbation}
\label{App:single_interface_minimalpot}
The minimum required interface perturbation to shift the real part depends on the band alignment of the semiconducting band relative to the Hubbard bands. Mathematically, it depends on the on-site potential $V$ as this shifts the band edge.
Now, one needs to distinguish four different cases: the semiconductor band edge is i) above of the upper Hubbard band; 
ii) between the band gap center $U/2$ and the upper Hubbard band; 
iii) above the lower Hubbard band but below the band gap center; 
iv) below the lower Hubbard band. 
In the first case, the minimum interface perturbation reads:
\bea	
&&4\Delta V_\mathrm{min,B}=\frac{U^2 Z}{4 T-Z (2 T_\mathbf{k}^\|+U-2 V)}+U-2 V \nn
&&-\frac{4 i T \sqrt{1-\frac{\left(\alpha+\beta\right)^2}{4 T^2 (Z (2 T_\mathbf{k}^\|+U-2 V)-4 T)^2}}}{Z}
\eea
with the abbreviations:
\begin{equation}
\begin{aligned}
    \alpha &= 8 T^2-4 T Z (T_\mathbf{k}^\|+U-2 V) \\
    \beta &= Z^2 \left(-2 V (T_\mathbf{k}^\|+U)+T_\mathbf{k}^\| U+2 V^2\right)
    \end{aligned}
\end{equation}
In the second case, we find: 
\bea
&&4 Z \Delta V_\mathrm{min, A_H}=\sqrt{4 T^2-4 T T_\mathbf{k}^\| Z+Z^2 \left((T_\mathbf{k}^\|)^2+U^2\right)} \nn
&&-4 i T \sqrt{1-\frac{\left(\alpha+2 T+Z (T_\mathbf{k}^\|+U-2 V)\right)^2}{16 T^2}} \nn
&&-2 T+Z (T_\mathbf{k}^\|+U-2 V)
\eea
with the abbreviation:
\begin{equation}
    \alpha =\sqrt{(2T-Z T_\mathbf{k}^\|)^2+Z^2U^2}
\end{equation}
as the minimum interface perturbation. In the third case, we get $\Delta V_\mathrm{min,B}$ again. In the fourth case we find:
\bea
&&4Z\Delta V_\mathrm{min,A_H}=-\sqrt{4 T^2-4 T T_\mathbf{k}^\| Z+(T_\mathbf{k}^\|)^2 Z^2+U^2 Z^2}\nn
&&-4 i T \sqrt{1-\frac{\left(-\alpha+2 T+Z (T_\mathbf{k}^\|+U-2 V)\right)^2}{16 T^2}}\nn
&&-2 T+T_\mathbf{k}^\| Z+U Z-2 V Z
\eea
with:
\begin{equation}
    \alpha =\sqrt{(2T-Z T_\mathbf{k}^\|)^2+Z^2U^2}
\end{equation}

\subsection{Single Interface at Mott-N\'eel Background}
Without loosing any generality, we use the mean-field state Eq.~\ref{eq:rho0MN}. This fixes the occupation number expectation values to $\langle \h  n_{\mu_A \uparrow}\rangle=1$, $\langle \h  n_{\mu_B \uparrow}\rangle=0$. Combining Eq.~\ref{difference2} with the \textit{ansatz} Eq.~\ref{eq:MNAnsatzSemi} and Eq.~\ref{eq:MNAnsatzMott}
yields four boundary conditions:
\begin{equation}
	\begin{aligned}
	(E-V-\Delta V)p_0^{0A}&=- \left[T_\mathbf{k}^\|p_0^{0B}+ \frac{T}{2Z}p_1^{1B}+\frac{T}{Z}p_{-1}^{0B}\right], \\
	Ep_1^{0A}&=- \left[T_\mathbf{k}^\|p_1^{1B}+ \frac{T}{Z}p_2^{1B}+\frac{2T}{Z}p_{0}^{0B}\right] ,\\
	(E-V-\Delta V)p_0^{0B}&=- \left[T_\mathbf{k}^\|p_0^{0A}+ \frac{T}{2Z}p_1^{0A}+\frac{T}{Z}p_{-1}^{0A}\right],\\
	(E-U)p_1^{1B}&=- \left[T_\mathbf{k}^\|p_1^{0A}+ \frac{T}{Z}p_2^{0A}+\frac{2T}{Z}p_{0}^{0A}\right].
\end{aligned}
\end{equation}
By using Eq.~\ref{bipartite-spinors} within the single regions these might be simplified to:
\begin{equation}
\label{eq:boundarycondMN}
	\begin{aligned}
	-\frac{Z}{T}\Delta V p_0^{0AS}+2p_1^{0BS}&=p_1^{1BM}, \\
	 2p_0^{0BS}&=p_0^{1BM},\\
	 -\frac{Z}{T}\Delta V p_0^{0BS}+2p_1^{0AS}&=p_1^{0AM},\\
2p_0^{0AS}&=p_0^{0AM}.
\end{aligned}
\end{equation}
The additional index $M$ or $S$ marks the Mott and semiconducting spinor, respectively.

\vfill
\bibliography{bib.bib}
\end{document}